\documentclass[letterpaper,11pt,fleqn]{article}
\usepackage{jheppub}

\usepackage{graphicx}
\usepackage{dcolumn}
\usepackage{bm}
\usepackage{epstopdf}
\usepackage{mathrsfs}
\usepackage{amssymb,amsfonts,latexsym}
\usepackage{amsmath,bbold}

\def\bs{\boldsymbol}
\def\del{\partial}

\newcommand{\eqn}[1]{Eq.~\eqref{#1}}
\long\def\comment#1{ }

\newcommand{\abar}{\bar{\alpha}}

\def\p{{\boldsymbol p}}
\def\bp{{\vec p}}
\def\bk{{\vec k}}
\def\bq{{\vec q}}

\def\q{{\bm q}}
\def\l{{\boldsymbol l}}
\def\k{{\boldsymbol k}}

\def\r{{\boldsymbol r}}

\def\u{{\boldsymbol u}}
\def\v{{\boldsymbol v}}

\def\Q{{\boldsymbol Q}}
\def\P{{\boldsymbol P}}

\newcommand{\beq}{\begin{eqnarray}}
\newcommand{\eeq}{\end{eqnarray}}
\newcommand{\be}{\begin{eqnarray*}}
\newcommand{\ee}{\end{eqnarray*}}

\newcommand{\rmd}{{\rm d}}
\newcommand{\rme}{{\rm e}}

\newcommand{\nn}{\nonumber\\ }

\title{\Large Probabilistic picture for medium-induced jet evolution}

\author{Jean-Paul Blaizot,}
\author{Fabio Dominguez,}
\author{Edmond Iancu,}
\author{and Yacine Mehtar-Tani}

\affiliation{%
Institut de Physique Th\'eorique, CEA Saclay, F-91191 Gif-sur-Yvette,
France}

\emailAdd{jean-paul.blaizot@cea.fr}
\emailAdd{fabio.dominguez@cea.fr}
\emailAdd{edmond.iancu@cea.fr}
\emailAdd{yacine.mehtar-tani@cea.fr}

\abstract{A high energy jet that propagates in a dense medium generates a cascade of partons that can be described as  a classical branching process. A simple generating functional for the probabilities to observe a given number of gluons at a given time is derived. This is used to obtain an evolution equation for the inclusive one-gluon distribution, that takes into account the dependence upon the energy and the transverse momentum of the observed gluon. A study of the explicit transverse momentum dependence of the splitting kernel leads us to identify large corrections to the jet quenching parameter $\hat q$.}

\arxivnumber{1307.xxxx}

\begin{document}
\maketitle

\section{Introduction}

The recent experimental results from  heavy ion experiments at RHIC and LHC provide a strong motivation for improving and  extending current theories of jet  propagation   in a dense QCD medium such as a quark-gluon plasma. Most noteworthy are those data that reveal  the jet inner structure
\cite{Aad:2010bu,Chatrchyan:2011sx,Chatrchyan:2012nia,Aad:2012vca,Chatrchyan:2013kwa}, and in particular show that much of the energy lost in the medium is in the form of low energy quanta emitted at large angles with respect to the jet axis. Such data call for the development of  new theoretical tools allowing us to explore  jet quenching phenomena beyond the energy loss from the leading particle, a subject that  has been thoroughly studied within the BDMPSZ framework during the last twenty years or so \cite{Baier:1996kr,Baier:1996sk,Baier:1998kq,Zakharov:1996fv,Zakharov:1997uu}. Further related developments are presented in \cite{Wiedemann:2000za,Gyulassy:2000er,Arnold:2002ja}. 

In order to understand how the QCD shower gets modified as the jet traverses a dense medium, one needs to study how  color coherence, and interference between subsequent emissions, are affected by the medium. These two effects play a major role in determining jet-structure in vacuum. Recent studies \cite{MehtarTani:2010ma,MehtarTani:2011tz,CasalderreySolana:2011rz,MehtarTani:2011gf,MehtarTani:2011jw,CasalderreySolana:2012ef} have shown how, in certain regimes,  color coherence can be destroyed by the presence of a dense medium, and how this may lead to the suppression of angular ordering (a characteristic feature of the vacuum QCD cascade), and the enhancement of large angle emissions. In a previous paper \cite{Blaizot:2012fh} we showed explicitly that the loss of color coherence occurs on a time scale comparable to that of the branching process, so that gluons that emerge from a splitting propagate independently of each other. The branching time, which is proportional to the square root of the energy of the emitted gluons, can be short for soft enough gluons, and multiple branching play an important role if the size of the medium is large enough: these multiple branching constitute the in-medium QCD cascade. 
 
 The goal of this paper is to complete the description of this cascade. It is organized as follows. In the next section we briefly recall the main results of Ref.~\cite{Blaizot:2012fh} concerning the properties of the medium induced gluon splitting and of transverse momentum broadening within the BDMPSZ framework. Then, in Section 3, we construct a generating functional for the probabilities to observe $n$ gluons in the cascade, at any given time. This is then used to derive the evolution equation for the inclusive one-gluon spectrum. This equation generalizes that studied in Ref.~\cite{Blaizot:2013hx} in that it takes into account the dependence of the distribution function on the transverse momentum of the produced gluon, as generated via collisions in the medium. 
 (The equation studied in \cite{Blaizot:2013hx} concerns only the energy distribution, that is, the integral of the one-gluon spectrum over the transverse momentum.) The kernel of this equation, however, is completely integrated over the transverse momenta and contains information on these transverse momenta only in an average way:
 this follows from the fact that the transverse momentum broadening acquired during
the branching processes can be neglected as compared to that accumulated via collisions in the medium
in between successive branchings. Thus, to the accuracy of interest, the splittings can be effectively treated
as being collinear.
 By trying to improve the description and take into  account more explicitly the transverse momentum dependence of the splitting kernel, we were led to identify large
 radiative corrections, which are formally infrared divergent, and are best interpreted as corrections to the transport coefficient $\hat q$, which is a measure of the transverse momentum square acquired by the jet parton in the medium, per unit length. This will be discussed in Section 4.
 In particular, we recover the double logarithmic correction to transverse momentum broadening that has been calculated recently \cite{Liou:2013qya}. Technical material is gathered in three Appendices. The first one complements results obtained in \cite{Blaizot:2012fh}, and gives an explicit expression for the splitting kernel in the harmonic approximation, with full dependence on transverse momenta. The contribution of the single scattering is emphasized. The second appendix  is devoted to the calculation of the double logarithmic contribution to $\hat q$. The third appendix presents an alternative form of the generating functional that may be more suitable for Monte-Carlo calculations.

\section{Basic elements}\label{sec:fact}

A large part of the material of this section is borrowed from Ref.~\cite{Blaizot:2012fh} to which we refer for more details. 
We consider energetic partons traversing a medium with which they exchange color and transverse momentum. This medium is modeled as a random color field, with the only non-trivial correlator (in light-cone gauge $A^+=0$)\footnote{To simplify the notation, the light-cone time variable $x^+=(x^0+x^3)/\sqrt{2}$ is labeled $t$ throughout the paper, and is referred to simply as `time' rather than light-cone time.  } given by
\beq\label{correlmed}
\left\langle A^-_a(\q,t)A^{*-}_b(\q',t')\right\rangle=\delta_{ab}n(t)\delta(t-t')(2\pi)^2\delta^{(2)}(\q-\q')\,\gamma(\q)\;,
\eeq
with $n$ the color charge density (which may depend on the light-cone time $t$), and we have assumed translational invariance in the transverse plane. Here, $\gamma(\q)\simeq  g^2/\q^4$, with $\q $ a vector in the transverse plane, is a correlator that accounts for the elastic collisions of the energetic patrons with the medium particles. The infrared behavior of $\gamma(\q)$ is controlled by the Debye screening mass $m_D$, which is the typical momentum exchanged in a collision with medium particles. 

Consider now the  process in which a gluon  is created inside the medium with momentum $\vec{p}_0\equiv(p_0^+,\p_0)$ at time $t_0$. In practice, we shall eventually chose $\bm{p}_0
= \bm{0}$, that is, we shall use the direction of motion of the leading particle  
to define the longitudinal direction with respect to which one measures polar angles and transverse momenta\footnote{However, in many  formulae below we shall keep $\p_0$ explicit, for more clarity, in particular in cases where one needs to emphasize differences between transverse momenta. The same remark applies to the initial time $t_0$ which can be chosen to be $t_0=0$.}.
For $t>t_0$, the gluon propagates through the medium and interacts with the latter. In leading order, that is, in the absence of splitting, all what happens to it is that it receives transverse momentum kicks in colliding with the medium constituents.  When it emerges from the medium, at time $t_L$\footnote{Note that $t_L=\sqrt{2}L$, with $L$ the length of the medium}, its momentum is $\vec{k}=(k^+,\k)$. Let us denote by $P_1(\vec{k};t_L,t_0)\rmd\Omega_k$ the probability to observe the gluon at time $t_L$ with its momentum in the phase-space element $\rmd\Omega_k$, given that, at time $t_0$  its momentum is in the phase-space element $\rmd\Omega_{p_0}$. Here, $d\Omega_k\equiv (2\pi)^{-3} \rmd^2\k\, d k^+/2k^+$ is the invariant phase-space element. The probability $P_1$ contains, quite generally, a delta function that expresses the conservation of the + component of the momentum  (that follows from the fact that the medium  is assumed  homogeneous in $x^-$), and it is convenient to write it as
\beq\label{P1}
P_1(\vec{k};t_L,t_0)=2p_0^+\, 2\pi\delta(k^+-p_0^+){\cal P}_1(\k;t_L,t_0).
\eeq
In leading order,  ${\cal P}_1(\k;t_L,t_0)$ can be identified with the probability for the gluon to acquire a transverse momentum $\k-\p_0$ from the medium during its propagation from time $t_0$ to time $t_L$, a quantity that we shall denote simply by ${\cal P}(\k-\p_0;t_L,t_0)$ throughout the paper\footnote{Remark on the notation: in ${\cal P}_1$ the dependence on $\p_0$ is kept implicit, while we leave $\p_0$ explicit in ${\cal P}$. This is because we shall need ${\cal P}$ also for differences of transverse momenta that do not necessarily involve $\p_0$.}. This probability ${\cal P}$  is well known, and it can be written as 
\beq\label{Pcoord}
{\cal P}(\k-\p_0;t_L,t_0)=\int \rmd^2\r \exp\left[-i(\k-\p_0)\cdot\r-\frac{N_c}{2}\int_{t_0}^{t_L}dt\,n(t)\sigma(\r)\right],
\eeq
with  $\sigma(\r)$ the `dipole cross section'
\beq\label{sigmadipole}
\sigma(\r)=2g^2\int\frac{\rmd^2\q}{(2\pi)^2}\left(1-\rme^{i\q\cdot\r}\right)\,\gamma(\q).
\eeq
Note that  $\sigma(r\to 0)\to 0$, a property commonly referred to as color transparency; this  ensures in particular that the probability ${\cal P}$ is properly normalized:
$\int_\k {\cal P}(\k-\p_0;t,t_0)\,=\,1$.
The Fourier transform of the dipole cross section reads
\beq\label{FTsigmadip}
\sigma(\l)=\int\rmd \r \,{\rm e}^{-i\l\cdot\r} \sigma(\r)=-2g^2\left[  \gamma(\l)-(2\pi)^2\delta(\l)\int_{\q}\gamma(\q) \right],
\eeq
and it obeys the properties (to be used later)
\beq\label{sigmaproperties}
\int_\l \sigma(\l)=0=\int_\l \l\,\sigma(\l).
\eeq
In the equations above,  we introduced a 
shorthand notation for the transverse momentum integrations: $\int_\q\equiv \int \rmd^2\q/(2\pi)^2$. This will be used  throughout the paper. 

\begin{figure}
\begin{center}
\includegraphics[height=2.7cm]{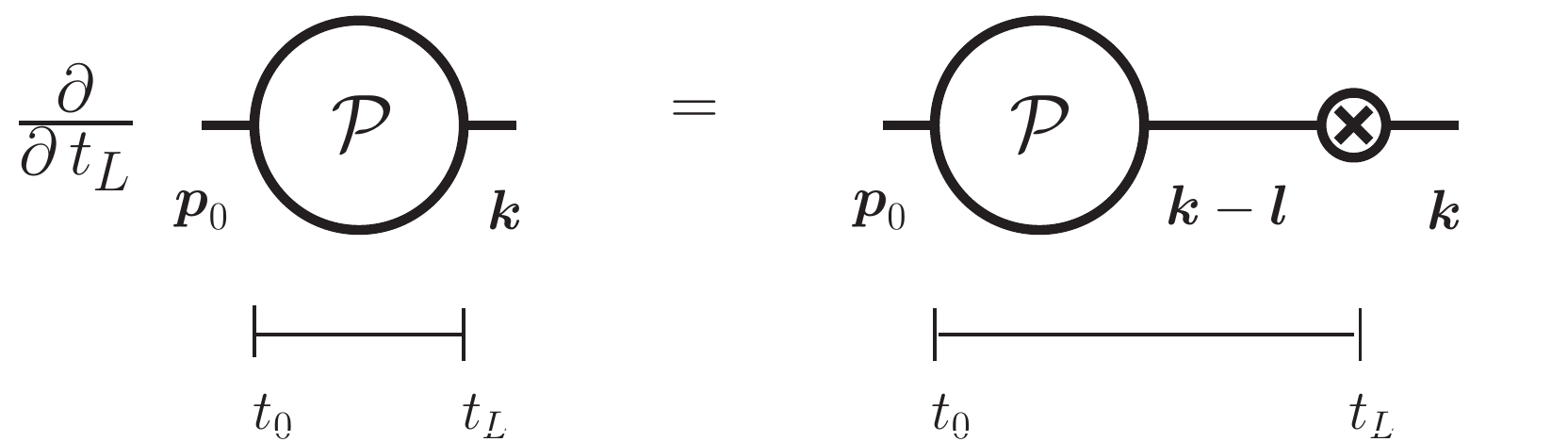}
\caption{Diagrammatic representation of the evolution of the broadening probability \eqref{difPC}.} \label{figure1}
\end{center}
\end{figure}

By taking the derivative of Eq.~(\ref{Pcoord}) with respect to $t_L$ (and setting $t_L=t$), one easily obtains
\beq\label{difPC}
\frac{\partial}{\partial t}{\cal P}(\k-\p_0;t,t_0)=\int_\l\,{\cal C}(\l,t)\,{\cal P}(\k-\p_0-\l;t,t_0)\,,
\eeq
with 
\beq\label{Vbar}
{\cal C}(\l,t)\equiv 4\pi\alpha_sN_cn(t)\left[\gamma(\l)-(2\pi)^2\delta^{(2)}(\l)\int_\q\gamma(\q)\right]=-\frac{1}{2} N_c n(t) \sigma(\l).
\eeq
This equation can be  simplified by taking into account the fact that the typical momentum
transferred in one collision is $|\l| \sim m_D$ and is much smaller than the
transverse momentum $|\k|\sim Q_s\equiv\sqrt{\hat qL}$ acquired by the gluon during its propagation over a distance
comparable to $L$, the size of the medium. Under such circumstances, Eq.~(\ref{difPC}) can be reduced to  the following Fokker-Planck equation:
\beq\label{difP}
\frac{\partial}{\partial t}{\cal P}(\k-\p_0;t,t_0)= 
\frac{1}{4}\frac{\del^2}{\del \k^2}\left[\hat q(t,\k^2)\,{\cal P}(\k-\p_0;t,t_0)\right]\,,
\eeq
with the jet quenching parameter $\hat q(t,\k^2)$, playing the role of a diffusion coefficient,  given by 
\beq\label{qhatsigmal}
\hat q(t,\k^2)=-{N_c n(t)}\int_{\l}\l^2\sigma(\l)=g^2 N_c n(t) \int_\l \l^2\gamma(\l)\approx 4\pi\alpha_s^2 N_c n(t)\ln\frac{\k^2}{m_D^2}.
\eeq
The above integral, which determines the value of $\hat q$, is logarithmically divergent. It is naturally cut-off at its lower end by the Debye mass, and  at its upper end   by  the momentum scale $\k$ at which ${\cal P}$ is evaluated.
Note that in deriving Eq.~(\ref{difP}) attention has been paid to this momentum dependence of $\hat q$. It can be verified in particular that, as written, the right hand side of the equation vanishes upon integration over $\k$, as it should. 

An alternative interpretation of  $\hat q$ is obtained by expanding the dipole cross section (\ref{sigmadipole}) to quadratic order in the dipole size. 
This yields 
 \beq\label{sigma}
N_cn(t)\sigma(\r)\simeq\frac{1}{2}\,\hat q(t,1/r^2)\,\r^2,
\eeq
where the inverse of the dipole size $r$ plays the role of ultraviolet cut-off. This expression, when used with a constant $\hat q$ (i.e., ignoring the dependence of $\hat q$ on the dipole size), is referred to as the `harmonic approximation'. Within this approximation, the diffusion equation (\ref{difP}) is easily solved. Assuming $n$, and hence $\hat q$, to be independent of $t$ for simplicity, one gets
\beq
{\cal P}(\k-\p_0;t,t_0)=\frac{4\pi}{\hat q (t-t_0)}{\rm e}^{-\frac{(\k-\p_0)^2}{\hat q (t-t_0)}}.
\eeq

The diffusion picture is valid in the regime dominated by multiple scattering, in which a large transverse momentum is achieved by the addition of many small momentum transfers over the propagation time $\Delta t=t-t_0$. This regime holds for $k^2\lesssim \hat q\Delta t$. Larger transverse momenta can be achieved, over comparable time scales, through a single hard scattering. The corresponding  expression for ${\cal P}$ is not given by the diffusion equation, but rather by using the first iteration of \eqn{difPC} or, equivalently, by expanding the exponential in \eqn{Pcoord} to linear
order in $\sigma$. Either way, one finds that in the regime where $k^2\gg \hat q (t_L-t_0)$:
\beq
{\cal P}(\k; t_L,t_0)\simeq\,\frac{16\,\pi^2\,\alpha_s^2\,N_c}{\k^4}\,\int_{t_0}^{t_L}dt\,n(t)\,.
\eeq

Let us now turn to the process of in-medium gluon branching which  was studied in detail in \cite{Blaizot:2012fh}. A major  assumption in that calculation is that the branching time $\tau_{_{\rm br}}\simeq\sqrt{ \omega_0/\hat q_0}$ is much shorter than the time $t_L-t_0$ spent by the partons in the medium, or equivalently $\omega_0\lesssim\omega_c=\hat q_0 t_L^2$, where $\omega_c$ is the maximum energy that can be taken away by an offspring gluon, i.e., 
$\tau_{_{\rm br}}(\omega_c)=t_L-t_0$. (We have set $\omega_0\equiv z(1-z)p_0^+$, and $\hat q_0\equiv \hat q f(z)$ with $f(z)=1-z(1-z)$ ; see also Appendix A and Eq.~(\ref{Kz2}) below.) 

Let $P_2(\vec{k}_a,\vec{k}_b;t_L,t_0)\rmd\Omega_{k_a}\rmd\Omega_{k_b} $ be the probability to observe two gluons at time $t_L$ in the phase space elements $ \rmd\Omega_{k_a}$ and $\rmd\Omega_{k_b}$, respectively, given that one gluon was present in the phase-space element $\rmd\Omega_{p_0}$ at time $t_0$. Similarly to what we did for $P_1$ in Eq.~(\ref{P1}), we separate the delta function that expresses the conservation of the + momentum and write
\beq\label{defP2}
P_2(\vec{k}_a,\vec{k}_b;t_L,t_0)&=& 2p_0^+\, 2\pi \delta(k_a^{+}+k_b^+-p_0^+)\, {\cal P}_2(\k_a,\k_b,z;t_L,t_0). 
\eeq
In Appendix A, it is recalled that when one drops all possible terms that are suppressed by at least one power of $\tau_{\rm br}/L$, one ends up with a simple formula for ${\cal P}_2$:
 \begin{align}\label{sigma1a2}
{\cal P}_2(\k_a,\k_b,z;t_L,t_0)&=2g^2z(1-z)
   \int_{t_0}^{t_L}\,\rmd t\,{\cal K}(z,p_0^+; t) \nn &\times  \int_{\q} \;{\cal P}(\k_a-z\q;t_L,t)\,
{\cal P}(\k_b-(1-z)\q;t_L,t)
 {\cal P}(\q-\p_0;t,t_0),
\end{align}
 where  ${\cal K}(z,p_0^+; t)$ is given by 
 \beq\label{Kz2}
{\cal K}(z,p_0^+;t)=
\frac{P_{gg}(z)}{2\pi}\sqrt{\frac{\hat q f(z) }{z(1-z)p_0^+}}=
\frac{P_{gg}(z)}{2\pi}\sqrt{\frac{\hat q_0 }{\omega_0}},\qquad f(z)=1-z+z^2,
\eeq
with 
\beq\label{Pgg}
P_{gg}(z)=N_c\,\frac{[1-z(1-z)]^2}{z(1-z)}=N_c\frac{[f(z)]^2}{z(1-z)}
\eeq
the leading--order Altarelli--Parisi gluon splitting function \cite{Altarelli:1977zs}.
We may interpret the integrand of Eq.~(\ref{sigma1a2}) as a product of probabilities: the probability ${\cal P}(\q-\p_0;t,t_0)$  for the initial gluon  to acquire transverse momentum $\q-\p_0$ in time $t-t_0$, the probability ${\cal K}(z,p_0^+; t)\rmd t$ for the gluon to split between times $t$ and $t+\rmd t$, into two gluons with energy fractions $z$ and $1-z$, and the probabilities for the offspring gluons to evolve to momenta $\k_a$ and $\k_b$, respectively ${\cal P}(\k_a-\p;t_L,t)$ and
${\cal P}(\k_b-\q+\p;t_L,t)$. 

Note  that the splitting described by Eq.~(\ref{sigma1a2}) is {\em collinear}~:
just after the splitting, the daughter gluons carries equal fractions, $z$ and respectively $1-z$, of both the longitudinal momentum $p_0^+$ and the transverse momentum $\q$ of their parent gluon. This is a consequence of the leading order approximation in which we ignore, in the factors ${\cal P}$, the small contribution to momentum broadening that may occur during the branching process (such small contributions  are taken into account in an average way in the  kernel). As a result, in the leading order approximation, transverse momentum is gained only in between the splittings, through momentum space diffusion.

Eq.~(\ref{sigma1a2}) will be at the basis of the classical branching process to be constructed
in the next section. However, in the last section of this paper we shall examine a more complete  version of the splitting kernel, which  keeps track of the transverse momentum that is acquired during  the branching process. This is obtained by relaxing some of the approximations leading to Eq.~(\ref{sigma1a2}), and involves corrections, a priori small since of order $\tau_{_{\rm br}}/L$, but which happen to be amplified by logarithmic divergences. As we shall see, these corrections are best interpreted as corrections to $\hat q$, or equivalently as corrections to the interaction between the partons of the cascade and the medium particles.

\section{Generating functional and inclusive one-gluon distribution}\label{sec:gf}

In this section, we introduce the generating functional that describes the in-medium gluon cascade, 
under the assumption that successive branchings can be treated as 
independent, in line with the results of Ref.~\cite{Blaizot:2012fh}. 
The generating functional follows simply from  iterating the
elementary $1\to 2$ splitting process whose properties are recalled in the previous section. From the generating functional, we derive the equation for the inclusive one-gluon distribution function and we analyze some of its properties. 

\subsection{Generalities}


We consider an  in-medium
parton shower initiated at (light-cone) time $t_0$ by a `leading parton' with 3--momentum 
${\vec p}_0\equiv (p^+_0,\bm{p}_0)$. 
The generating functional ${\cal Z}_{p_0}\,[t,t_0|u]$, with $t_0\le t\le t_L$,   is defined as
 \beq
{\cal Z}_{p_0}[t,t_0|u]=\sum_{n=1}^\infty\frac{1}{n!}\int\left( \prod_{i=1}^n \,\rmd\Omega_i\right)
\, P_n(\bk_1,\cdots,\bk_n;t,t_0)\,u(\bk_1)\cdots u(\bk_n)\,
\eeq
where $u\equiv u(\bk)$ is a generic function of $\bk$ and $P_n(\bk_1,\cdots,\bk_n;t,t_0)$ is the probability density 
to find at time $t$ exactly
$n$ gluons with momenta $\bk_1,\cdots,\bk_n$ such that  $k^+_1+\cdots+k^+_n= p_0^+$ (recall that the + component of the momentum is conserved during the branchings). The function $P_n(\bk_1,\cdots,\bk_n;t,t_0)$ is totally
symmetric under the permutations of the $n$ variables $\bk_1,\cdots,\bk_n$. Leading order expressions for the probabilities $P_1$ and $P_2$ have been given in the previous section. Note that,  $ {\cal Z}_{p_0}[t,t_0 | u=1]=1$, which reflects the normalization of the probabilities,  while obviously $ {\cal Z}_{p_0}[t,t_0| u=0]=0$.

By taking  the $n$th functional derivative
of $ {\cal Z}_{p_0}[t,t_0 | u]$ evaluated at $u=0$, one recovers  $P_n(\bk_1,\cdots,\bk_n;t,t_0)$:
 \beq\label{n-prob}
P_n(\bk_1,\cdots,\bk_n;t,t_0)=\left.\left[\prod_{i=1}^n \, (2\pi)^3\,2k^+_i\,
\frac{\delta}{\delta u(\bk_i)}\right] {\cal Z}_{p_0}[t,t_0 | u]\right|_{u=0}\,,
\eeq
with the usual definition
\beq\label{var-u}
\frac{\delta u(\bk)}{\delta u(\bq)}=\delta^{(3)}(\bk-\bq)\equiv 
\delta(k^+-q^+)\,\delta^{(2)}(\k-\q) \,.
\eeq
We shall be mostly concerned with inclusive probabilities, that is by the probabilities to observe at time $t$, $n$ gluons with specified momenta, irrespective of whether other gluons are produced or not. Such probabilities are obtained  by taking the $n$-th functional derivative of ${\cal Z}_{p_0}[t,t_0|u]$
and then letting $u=1$. 

\subsection{Evolution equations for the generating functional}

Two formulations can be considered for the evolution of the generating functional. 
We consider in this section the 
`forward', formulation, where an additional splitting is allowed to occur at the latest time $t$ of the cascade development. In Appendix C, another formulation  is presented, where one focuses instead on a splitting taking place at the beginning of the cascade. Both formulations are equivalent but lead to different equations (see e.g. Ref.~\cite{Cvitanovic:1980ru} for a general discussion).

Let us then consider the initial state of the cascade, where a single gluon is present at time $t=t_0$. 
For $t=t_0$, all probabilities vanish except $P_1({\vec k};t_0,t_0)=2 k^+ (2\pi)^3 \delta(\vec{k}-\vec{p}_0)$, so that the generating functional reduces to 
 \beq\label{Zt0}
 {\cal Z}_{p_0}[t_0,t_0|u]\,=\,u(\bp_0)\,.\eeq
 During the infinitesimal time step $t_0\to t_0+\rmd t$,  two physical effects can occur: momentum broadening and splitting. Only the variations with time of $P_1$ and $P_2$ contribute: $P_2$ changes because a splitting can occur during the time $\rmd t$; $P_1$ changes for two reasons: first the collisions change the transverse momentum, second, probability conservation forces $P_1$ to decrease as  $P_2$ increases. Thus, at time $t_0+\rmd t$,  
 the generating functional reads
\beq
{\cal Z}_{p_0}[t_0+\rmd t,t_0|u]&=&\int \rmd\Omega_{k}\, P_1(\vec{k};t_0+\rmd t,t_0) \,u(\vec{k})\nn &+&\frac{1}{2}\int \rmd\Omega_{k_1} \rmd\Omega_{k_2}
\, P_2(\bk_1,\bk_2;\bp\,;t_0+\rmd t,t_0)\,u(\bk_1)u(\bk_2),
\eeq
where $P_1$ contains, besides the leading order contribution, a contribution of order $\alpha_s$, for the reason just mentioned.
 
 The leading order variation of $P_1$, which corresponds to momentum broadening, is easily deduced from  Eq.~(\ref{difPC}). The order $\alpha_s$ correction will be inferred from the conservation of probability. For $P_2$ we use the definition (\ref{defP2}) to write
 \beq
&& \frac{1}{2} \int\rmd\Omega_1\rmd\Omega_2 \,P_2(\vec{k}_1,\vec{k}_2;t,t_0) \,u(\vec{k}_1) u(\vec{k}_2)\nn
 &&=\frac{1}{4\pi} \int_0^1 \frac{\rmd z}{2z(1-z)} \int_{\k_1,\k_2} \, {\cal P}_2(\k_1,\k_2,z;t,t_0) \;u(zp_0^+,\k_1)\, u((1-z)p_0^+,\k_2),
 \eeq
 where we have used 
 \beq
 k_1^+=zp_0^+,\quad k_2^+=(1-z)p_0^+,\quad \rmd\Omega_1\rmd\Omega_2=\frac{1}{(2\pi)^2} \frac{\rmd z \rmd p_0^+}{4z(1-z) p_0^+}.
 \eeq
 Next, by taking a derivative w.r.t. $t_L$ on (\ref{sigma1a2}), and relabeling $t_L\to t$, $\k_a\to
 \k_1$, and $\k_b\to k_2$, one deduces 
\begin{align}
\label{diffP2L}
\del_{t}{\cal P}_2(\k_1,\k_2,z;t,t_0)=&2g^2z(1-z)
{\cal K}(z,p_0^+; t)\nn 
&\times\int_{\q} (2\pi)^4\delta^{(2)}(\k_1-z\q)\delta^{(2)}(\k_2-(1-z)\q)
{\cal P}(\q-\p_0;t,t_0), 
 \end{align}
and hence [recall that ${\cal P}(\q-\p_0;t_0,t_0)=(2\pi)^2\delta^{(2)}(\q-\p_0)$]
 \begin{align}\label{diffP2Lb}
{\cal P}_2(\k_a,\k_b,z;t_0+\rmd t,t_0)=2g^2z(1-z)\,
{\cal K}(z,p_0^+; t_0)\rmd t\ (2\pi)^4\delta^{(2)}(\k_1-z\p_0)\delta^{(2)}(\k_2-(1-z)\p_0). 
 \end{align}
 Combining these results, on can rewrite ${\cal Z}_{p_0}[t_0+\rmd t,t_0 | u]$ as follows 
 \begin{align}
\label{GFdiff-fwd0}
\left.\frac{\partial}{\partial t} {\cal Z}_{p_0}[t,t_0 | u]\right|_{t=t_0}=
&\int_{\l}\,{\cal C}(\l,t_0)\,u(p_0^+,\p_0+\l)\nn
&+\alpha_s\int_0^1\rmd
z\, {\cal K}(z,p_0^+;t)\Big[ u(zp_0^+,z\p_0)\,u((1-z)p_0^+,(1-z)\p_0)-u(\vec{p}_0)\Big].
\end{align}
Note that the last term, proportional to $u(\vec{p}_0)$ is here to ensure that the probability is conserved during the evolution\footnote{This term proportional to $u(\vec{p}_0)$ stands for the order $\alpha_s$ corrections to $P_1$ that we mentioned above.}: if one sets $u=1$, then all terms in the right-hand-side of Eq.~(\ref{GFdiff-fwd0}) vanish (recall that $\int_\l {\cal C}(\l)=0$).  

Equation (\ref{GFdiff-fwd0}) is easily extended to a full  evolution equation for the generating functional 
This  evolution equation for ${\cal Z}$ reads
\beq\label{GFdiff-fwd}
&&\frac{\partial}{\partial t_L} {\cal Z}_{p_0}[t_L,t_0| u]-\int
\frac{\rmd q^+}{2\pi}\int_{\q} \int_{\l}u(q^+,\q+\l)\,{\cal C}(\l,t_L)\,\frac{\delta }{\delta u(\vec{q})}\, {\cal Z}_{p_0}[t_L,t_0 | u]
\nn
&=&\alpha_s\int_0^1\rmd
z\, \int
\frac{\rmd q^+}{2\pi}\int_{\q} 
{\cal K}(z,q^+;t)\Big[ u(zq^+,z\q)\,u((1-z)q^+,(1-z)\q)-u(\vec{q})
\Big]\frac{\delta }{\delta u(\vec{q})} {\cal Z}_{p_0}[t_L | u]\,.\nn
\eeq
This formula has a simple interpretation. 
The effect of the functional derivative $\delta/\delta u(\bq)$  is to select a
gluon with momentum $\bq$ from the gluon cascade at time $t_L$. Then one calculates the evolution of this particular gluon by repeating the infinitesimal time step discussed before. The second term in the first line accounts for the collision of the gluon with the medium which, at time $t_L$,  turns  its momentum $\q$ into $\q+\l$. The second line contains the probability for this gluon to split
(via a collinear splitting), or not, in the time step
$t_L\to t_L+\rmd t_L$. In Appendix C, an equivalent equation is provided (cf.  Eq.~(\ref{GFdiff})), in which the generating functional is differentiated with respect to $t_0$. 
 
At this point, it is worth recalling that the equations above are somewhat formal since the integrals over the spliting
fraction $z$ develop endpoint singularities at $z=0$ and $z=1$. To see that more precisely, let us consider the evolution equation for  the  
probability ${\cal P}_1(\k; t_L,t_0)$. By using the definition  \eqref{n-prob} together with  the evolution equation \eqref{GFdiff-fwd}, one easily finds
\begin{align}\label{P1-evol}
\frac{\del}{\del t_L}{\cal P}_1(\k; t_L,t_0) - \int_{\l} {\cal C}(\l,t_L)
\,{\cal P}_1(\k-\l; t_L,t_0)
=-\alpha_s\int_0^1 \rmd z\,
{\cal K}(z,p^+_0;t_L)  \,{\cal P}_1(\k; t_L,t_0)\,,
\end{align}
where the r.h.s. originates from the `loss' term in \eqn{GFdiff-fwd}
and describes the reduction of the one-gluon probability due to branching. This equation
is easily solved by writing
\beq\label{P1-sol}
{\cal P}_1(\k; t_L,t_0)\,=\,& \Delta(p^+_0;t_L, t_0)\,{\cal P}(\k-\p_0; t_L,t_0).
\eeq
One then easily finds
\beq\label{Delta}
\Delta(p^+_0;t_L, t_0)\,=\,&
\exp\left[-\alpha_s\int_{t_0}^{t_L} \rmd t\int_0^1 \rmd z\, {\cal K}(z,p^+_0;t)\right].
\eeq
The physical meaning of \eqn{P1-sol} is transparent: ${\cal P}_1(\k; t_L,t_0)$ appears as the product of the probability 
 ${\cal P}(\k-\p_0; t_L,t_0)$ for the initial gluon (with momentum $\vec{p}_0$) to acquire transverse momentum $\k-\p_0$ via collision with the medium, multiplied by the `survival probability'   $\Delta(p^+_0;t_L, t_0)$ (aka the `Sudakov factor'), that is the probability for this gluon not to branch between $t_0$ and $t_L$. As it stands, this survival probability vanishes because of the endpoint singularities of the kernel ${\cal K}(z,p^+_0;t)$ at $z=0$ and $z=1$. 
A cut-off needs to be introduced,  which defines the `resolution', i.e., the energy below which gluons cannot be resolved anymore. 
The following identities ($\epsilon\to 0$), that result from the symmetry of the kernel in the substitution $z\to 1-z$, 
 \beq\label{endpointssing}
  \int_\epsilon^{1-\epsilon} \rmd z \,{\cal K}(z) \,=\,2 \int_0^{1-\epsilon} \rmd z \,z\,{\cal K}(z) 
  \,=\,2 \int_\epsilon^{1} \rmd z \,(1-z)\,{\cal K}(z) 
 \,,\eeq
allow us to concentrate  on one of the two endpoint singularities,
say that at $z=0$. One then easily estimates
 (with $\Delta t=t_L - t_0$)
\beq\label{Delta1}
\Delta(\epsilon p^+_0; \Delta t)\simeq\,
\exp\left[-2\abar\Delta t \sqrt{\frac{\hat q}{\epsilon p^+_0}}\,
\right],
\eeq
where $\abar\equiv \alpha_s N_c/\pi$. It follows in particular that  the typical time between two successive branchings
(the value of $\Delta t$ for which the exponent becomes of order one) is given by 
 \beq\label{trad}
 \Delta t_{\rm rad}(\epsilon p^+)\,\simeq\,\frac{1}{2\abar}\, \sqrt{\frac{\epsilon p^+}{\hat q}}\,\sim\,
 \frac{1}{\abar}\, \tau_{_{\rm br}}(\epsilon p^+)\,,\eeq
for a gluon within the cascade with generic energy $p^+$. 
In order for successive branchings to proceed independently from each other, we need $ \Delta t_{\rm rad}$ to be 
significantly larger than the duration $\tau_{_{\rm br}}(p^+)$ of the branching giving birth to the
$p^+$ gluon, which implies\footnote{Note that this is not a very restrictive condition for the
medium--induced cascade. Indeed, as demonstrated in Ref. \cite{Blaizot:2013hx}, 
the branchings of the soft gluons are mostly `democratic' as soon as $p^+\lesssim \abar^2
\omega_c$.} $\epsilon > \abar^2$. Such a cut--off must be included to give a meaning
to the generating functional. Note however that physical (inclusive) distributions remain finite
in the limit $\epsilon\to 0$, as we shall shortly verify,  hence they are only
weakly sensitive to the precise value of this cutoff, so long as it is small enough. 

\subsection{Evolution equation for the one-gluon distribution}
\label{sec:spec}

We turn now to a specific study of the inclusive one-gluon distribution. To simplify the notation, we shall omit the explicit dependence on the initial momentum ${\vec p}_0\equiv (p^+_0,\bm{0})$, as well as on $t_0$, and denote the gluon distribution simply by $D(x,\k,t)$, with $x=k^+/p^+_0$. We shall also assume from now on that the density $n$ is independent of time, and therefore so is  ${\cal K}$.

The inclusive one-gluon distribution is given by 
\beq\label{gluon-density}
D(x,\k,t)=k^+\frac{\rmd N}{\rmd k^+\rmd^2\k}&\equiv& k^+\left\langle\sum_{n=1}^\infty \sum_{j=1}^n
\delta^{(3)}(\bk_j-\bk)\right\rangle\nn
&=&   \frac{1}{2(2\pi)^3} \left\{  P_1(\vec{k};t,t_0) +\sum_{n=2}^\infty\frac{1}{n!}\int \prod_{i=1}^{n-1} \,\rmd\Omega_i 
\, n\, P_n(\bk, \bk_1,\cdots,\bk_{n-1};t,t_0)\right\}  \nn
   &=&\left.  k^+ \frac{\delta {\cal Z}_{p_0}[t,t_0|u]}{\delta u(\bk)}\right|_{u=1}\,.
\eeq 
\begin{figure}[h]
\includegraphics[height=3.2cm]{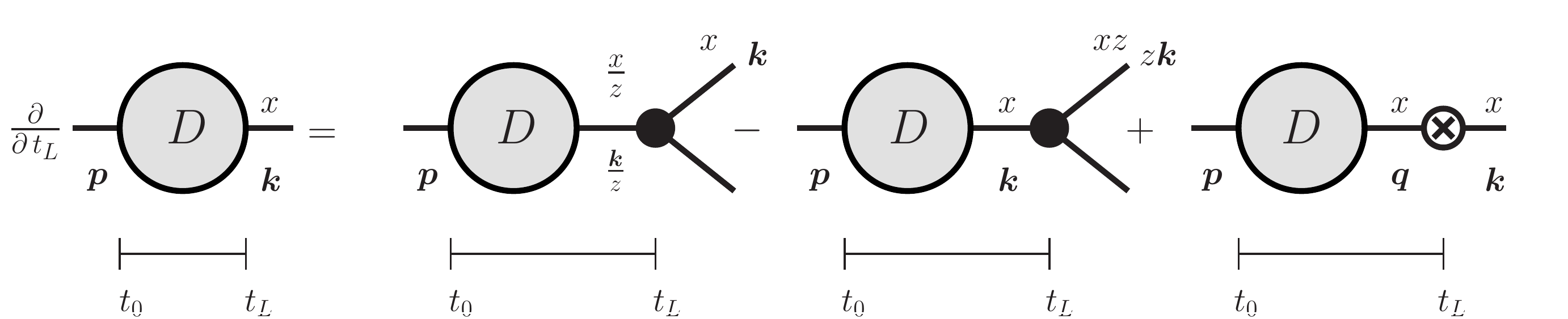}
\caption{Diagrammatic representation of the forward evolution for the inclusive gluon distribution.} \label{figure4}
\end{figure}

According to this formula, 
the evolution equation obeyed by $D(x,\k,t)$ can be obtained by taking a functional derivative 
$\delta/\delta u(\bk)$ of Eq. (\ref{GFdiff-fwd}), and then setting $u=1$ (note that only the explicit 
factors of $u$ in Eq.~(\ref{GFdiff-fwd}) contribute in this operation). We thus find
\begin{align}\label{Dkt10}
\frac{\partial}{\partial t}D(x,\k,t)&=\int_{\l}\,
{\cal C}(\l,t)D\left(x,\k-\l,t\right)\nn
&+\alpha_s\int_0^1\rmd
z \bigg[\frac{2}{z^2}{\cal K}\left(z,\frac{x}{z}p^+_0;t\right)
D\left(\frac{x}{z},\frac{\k}{z},t\right)-{\cal K}\left(z,xp^+_0;t\right)D\left(x,\k,t\right)\bigg].
\end{align}

This equation, which represents an important result of this paper,  is illustrated in Fig.~\ref{figure4}. The first term in its r.h.s. describes  transverse momentum broadening via medium rescattering
in between successive branchings, and leads to diffusion in momentum space. The two terms within the square brackets in the second line  of \eqn{Dkt10}  can be viewed respectively as 
a `gain term' and  a `loss term' associated with
one branching. The `gain term' describes the production of a new gluon with
energy fraction $x$ and transverse momentum $\k$ via the decay of an ancestor gluon
having energy fraction $x'=x/z>x$ and transverse momentum $\k/z$. (Note that the
condition $x < x' < 1$ implies $1 > z > x$ for the respective integral over $z$.) The `loss term' describes 
the disappearance of a gluon with energy fraction $x$ via the decay $(x,\k)\to (zx,z\k)\, ((1-z)x,(1-z)\k)$, 
with $0 < z <1$. Equation (\ref{Dkt10}) thus describes the interplay between collinear splittings (cf. Eq.~(\ref{sigma1a2})) and diffusion in momentum space in the development of the in-medium cascade.

\subsection{Energy distribution}

By integrating \eqn{Dkt10} over the transverse momentum $\k$, one finds a simplified
equation describing the evolution of the energy distribution alone:
 \begin{align}\label{Dxeq}
\frac{\partial}{\partial t}D(x,t)&=\alpha_s\int_0^1\rmd
z\Big[2{\cal K}\left(z,\frac{x}{z}p^+_0,t\right)
D\left(\frac{x}{z},t\right)
-{\cal K}\left(z,xp^+_0,t\right)D\left(x,t\right)\Big],
 \end{align}
 where we have set $D(x,t)\equiv \int_\k D(x,\k,t)$.
 Since the kernel is independent of time, the gluon distribution depends upon $t$ and $t_0$ only via their difference $t-t_0$ and it is convenient to rescale the time variable and the emission kernel in such a
way as to construct dimensionless quantities. Namely, we define
\beq\label{hatK}
\tau \equiv\,\frac{\alpha_s N_c}{\pi}\sqrt{\frac{\hat q}{p^+_0}}\,(t-t_0)\,,\qquad
\hat{\cal K}(z)\equiv \frac{2\pi}{N_c}\sqrt{\frac{p^+_0}{\hat q}}\,{\cal K}(z,p^+_0)\,=\,\frac{[1-z(1-z)]^{5/2}}
{[z(1-z)]^{3/2}}\,.
\eeq
Using also the property (cf. \eqn{Kz2}) 
\beq
{\cal K}\left(z,\frac{x}{z}p^+_0\right)\,=\,\sqrt{\frac{z}{x}}\,{\cal K}(z,p^+_0),\eeq
 as well as the identities Eq.~(\ref{endpointssing}), one can put the evolution equation \eqref{Dxeq} in the form
 \beq\label{Dfin}
\frac{\partial}{\partial\tau} D(x,\tau)=\int \rmd z \,
\hat{\cal K}(z)\left[\sqrt{\frac{z}{x}}D\left(\frac{x}{z},\tau\right)-\frac{z}{\sqrt{x}}D(x,\tau)\right],
\eeq
which is the equation\footnote{Eq.~(\ref{Dfin}) has been heuristically proposed in Refs.~\cite{Baier:2000sb,Jeon:2003gi} and later implemented in the MARTINI event generator \cite{Schenke:2009gb}. Recently \cite{Blaizot:2013hx}, a complete analytical study of this equation has been achieved showing its relevance in explaining the energy flow at large angles, via soft particles, responsible for dijet asymmetry \cite{Chatrchyan:2011sx}.}
 that has been studied in Ref. \cite{Blaizot:2013hx}.

Note that the singularities of   the kernel $\hat{\cal K}(z)$ at $z=0$ and $z=1$ are here harmless, since they exactly cancel: the integral over
$z$ in the `gain' term is restricted to $z > x$, while that in the `loss' term involves an additional
factor of $z$, which ensures convergence as $z\to 0$. When $z\to 1$, the `gain' and `loss' terms would
separately be singular, yet the respective divergences cancel in their sum, provided the
spectrum $D(x,\tau)$ is a regular function of $x$ for $x<1$. Accordingly,
\eqn{Dfin} is well defined as written, and the same applies to Eq.~(\ref{Dkt10}).

\section{Radiative corrections to $\hat q$}
\label{sec:hatq}

As recalled in the Appendix A, several approximations are involved in the derivation of the splitting kernel. Among those are approximations in which one ignores small momenta in the propagators ${\cal P}$, thereby allowing us to integrate the kernel over those particular momenta. Such approximations are in line with the leading order of our approximation scheme. However, in integrating the kernel over the various transverse momenta on which it may depend, one eliminates potentially interesting physics: we have seen in particular that in the leading order of our approximation scheme,  the gluon splitting is strictly collinear, with all transverse momenta arising from collisions with medium constituents in between the splittings. Clearly, one may wish to go beyond this simplified picture. In fact, the detailed calculations reported in Appendix A allow us to to go beyond the leading order approximation and explore the consequences of keeping some of these momenta in the factors ${\cal P}$ attached to a gluon splitting. Since these momenta are small, their effects can be well captured by a Taylor expansion, so that the entire corrections to the leading calculation presented so far appear as integral moments of the kernel, which may become large (in fact they are logarithmically divergent) for splittings that involve very soft gluons. As we shall see, these corrections are in fact better interpreted as corrections to the transport coefficient $\hat q$, or equivalently as corrections to the interaction between the partons of the cascade and the medium particles. Clearly, with the present discussion we cannot claim of having a systematic control of these corrections. However, we shall be able to identify the main correction to $\hat q$, in line with a recent study of transverse momentum broadening \cite{Liou:2013qya}.

 It is recalled  in Appendix A that the most general expression for the splitting probability that is 
compatible with the minimal set of approximations [referred to 1) and 2) in the Appendix] is given by
\begin{align}\label{sigma1aQ}
{\cal P}_2(\k_a,\k_b,z;t_L,t_0)&=2g^2z(1-z)
   \int_{t_0}^{t_L}\,\rmd t\,\int_{\q,\Q,\l}\,{\cal K}(\Q,\l,z,p_0^+; t) \nn &\times  {\cal P}(\k_a-\p;t_L,t)\,
{\cal P}(\k_b-(\q+\l-\p);t_L,t)
 {\cal P}(\q;t,t_0),
\end{align}
where $\Q=\p-z(\q+\l)$, with $\q$  the momentum of the gluon before splitting,  $\p$ that of the offspring that carries $zq^+$, and $\l$ is the transverse momentum acquired during the branching process. 
\begin{figure}[htbp]
\begin{center}
\includegraphics[width=7.5cm]{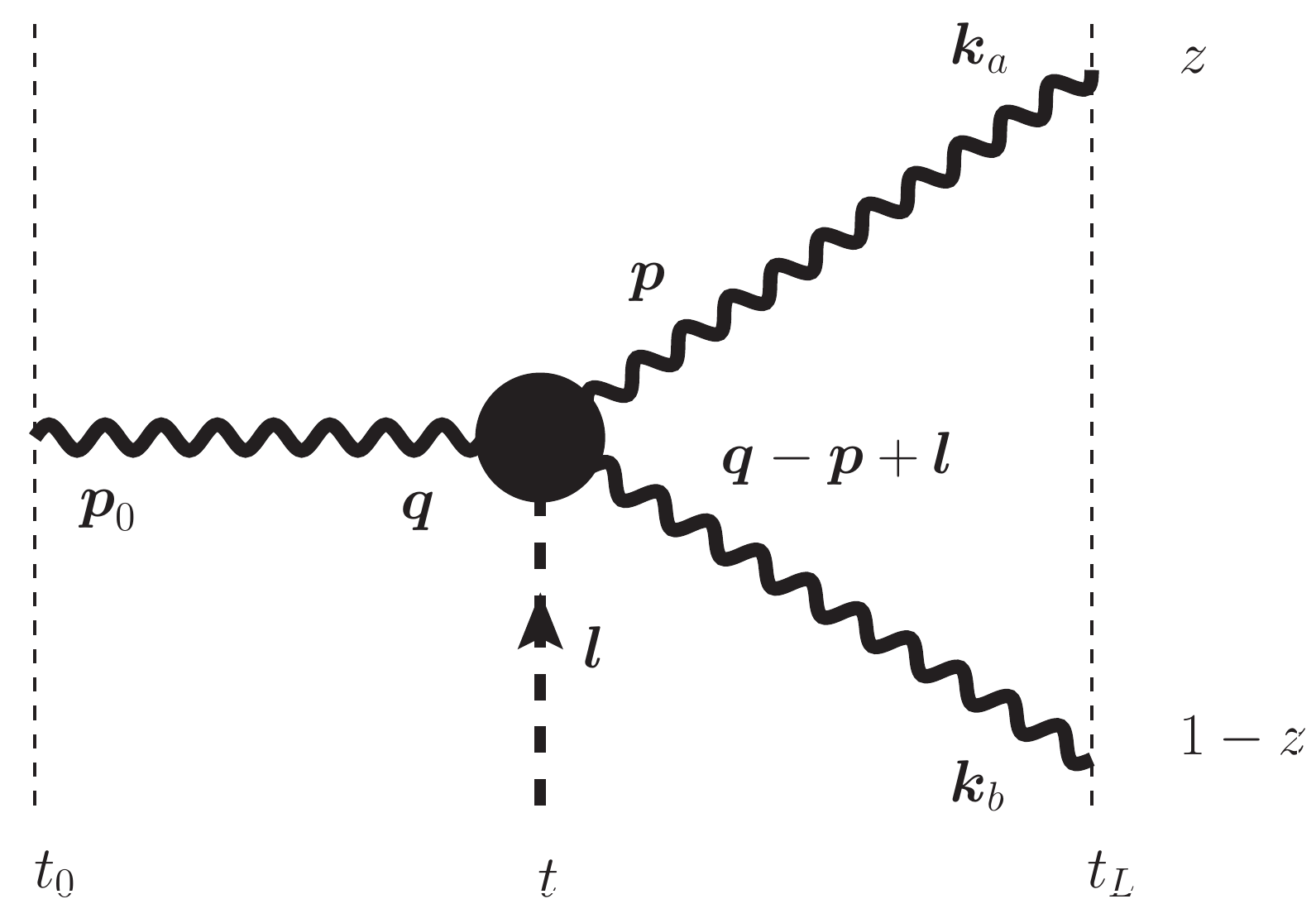}
\caption{\sl Graphical illustration of the equation (\ref{Dkt}).
The thick wavy lines represent the probability ${\cal P}$ for transverse
momentum broadening,  the black dot is the splitting kernel
${\cal K}$. }
\label{fat_vertex2}
\end{center}
\end{figure}
The complete expression of the splitting kernel 
${\cal K}(\Q,\l,z,p_0^+; t)$ is given in Appendix A, in terms of an integral representation obtained in the harmonic approximation (see Eq.~(\ref{kernel-harmonic})).  
Note that, in contrast to the fully integrated kernel in \eqn{Kz2}, the non integrated one is not positive definite
anymore. (This is already obvious on the partially integrated one, Eq.~(\ref{Klza}), although we may argue that this particular kernel becomes 
negative only in a momentum region where it is dwarfed by the exponential.) Yet, even though strictly speaking one loses their probabilistic interpretation, the manipulations of the previous section can be formally repeated in order to obtain the evolution equation for the inclusive one-gluon distribution corresponding to a more general splitting kernel. This equation reads
 \begin{align}\label{Dkt}
&\frac{\partial}{\partial t}D(x,\k,t)=\int_{\l}\,
{\cal C}(\l,t)D\left(x,\k-\l,t\right)\nn
&+\alpha_s\int_0^1\rmd
z\int_{\q,\l} \bigg[2{\cal K}\left(\Q,\l,z,\frac{x}{z}p^+_0\right)
D\left(\frac{x}{z},\q,t\right)-{\cal K}\left(\q,\l,z,xp^+_0\right)D\left(x,\k-\l,t\right)\bigg]\,,
\end{align}
where $\Q\equiv\k - z(\q+\l)$.  In the following, we shall use the fact that $\Q$ and $\l$ are generically small compared to $\k$ in order to simplify this equation. The fact that $\l$ is small is obvious from its interpretation as the momentum broadening acquired during the branching process. That $\Q$ is also small may be inferred from the explicit expression (\ref{Klza}) of the splitting kernel after integration over $\l$: this expression shows that the kernel 
which enters the `gain' term in \eqn{Dkt} is peaked around $|\Q|^2\sim k_{_{\rm br}}^2
\equiv\sqrt{\omega_0\hat q_0}$, with $\omega_0=(1-z)xp_0^+$.  The strategy that we shall follow then is the same as that we used in order to reduce Eq.~(\ref{difPC})  to the  diffusion equation (\ref{difP}), which involves essentially an expansion around the large momentum $\k$ of the followed gluon.

\subsection{The double logarithmic correction to $\hat q$}

In order to perform this expansion in powers of the small momenta $\Q$ and $\l$, it is convenient to change variables in the r.h.s. of \eqn{Dkt}, 
in such a way that these momenta  become the independent integration variables:
 \begin{align}\label{Dkt1a}
\frac{\partial}{\partial t_L}D(x,\k,t_L)&=\alpha_s\int_0^1\rmd
z\int_{\Q,\l}\,\bigg[\frac{2}{z^2}\,
{\cal K}\left(\Q,\l,z,\frac{x}{z}p^+_0\right)
D\left(\frac{x}{z},(\k-\Q -z\l)/{z},t_L\right)\nn
&-{\cal K}\left(\Q,\l,z,xp^+_0\right)D\left(x,\k-\l,t_L\right)\bigg]-\int_{\l}\,{\cal C}(\l)D\left(x,\k-\l,t_L\right)\,.
\end{align}
We can now expand the gluon distributions around $\k$. One gets, for the first term of Eq.~(\ref{Dkt1a}),
\beq\label{D-exp1}
D\left(\frac{x}{z},\frac{\k-\tilde\Q}{z}\right)=  D\left(\frac{x}{z},\frac{\k}{z}\right)-\tilde\Q\cdot\frac{\del}{\del \k}
D\left(\frac{x}{z},\frac{\k}{z}\right)
+\frac{1}{2!}\, \tilde Q^i \tilde Q^j\frac{\del}{\del k_i}\frac{\del}{\del k_j}D\left(\frac{x}{z},\frac{\k}{z}\right)+\cdots
 \eeq 
 where we have set $\tilde\Q\equiv \Q +z\l$. One expands similarly $D\left(x, \k-\l\right)$. 
It is easy to see that the leading terms will reproduce Eq.~(\ref{Dkt10}). The linear terms will vanish upon angular integration. Remain the quadratic terms, whose contribution can be cast in the form of the diffusion term, thereby exhibiting a correction $\delta\hat q$ to the jet quenching parameter. 
For consistency, we shall also simplify the collision term by using the diffusion approximation.

The evolution equation obtained after this expansion to quadratic order reads
 \begin{align}\label{Dkt-diff}
\frac{\partial}{\partial t_L}D(x,\k,t_L)&\,=\,
\alpha_s\int_0^1\rmd z\,\bigg[\frac{2}{z^2}\,{\cal K}\left(z,\frac{x}{z}p^+_0\right)
D\left(\frac{x}{z},\frac{\k}{z},t_L\right)-
{\cal K}\left(z,xp^+_0\right)D\left(x,\k,t_L\right)\bigg]\nn
&\quad +\frac{1}{4}\,\left(\frac{\del}{\del\k}\right)^2 \big[\hat q(\k^2)\,D\left(x,\k,t_L\right)\big]\nn
&\quad +\frac{1}{4}\, \left(\frac{\del}{\del\k}\right)^2
\int_x^1 \rmd z\,\frac{\rmd\delta\hat q (z,xp^+_0, \k^2)}{\rmd 
z}\, D\left(\frac{x}{z},\frac{\k}{z},t_L\right)\,,
\end{align}
where the first two lines are recognized as the leading--order transport equation, 
\eqn{Dkt10}, and in the last term we have set
 \begin{align}\label{qhat-rad}
\frac{\rmd\delta\hat q (z,xp^+_0, \k^2)}{\rmd z} 
 &\equiv \frac{2\alpha_s}{z^2}\int_{\Q,\l}\,
 (\Q+z\l)^2\,
{\cal K}\left(\Q,\l,z,\frac{x}{z}p^+_0\right)\nn
& - \alpha_s \delta(1-z)\int_0^1 \rmd z' 
\int_{\Q,\l}\,\l^2\,{\cal K}(\Q,\l,z',xp^+_0)\,.
 \end{align}
The $\k^2$--dependence in \eqn{qhat-rad} comes via the upper cutoff $\sim k$
in the integrals over $\Q$ and $\l$, which is kept implicit (see the discussion after Eq.~(\ref{qhatsigmal}), and Eq.~(\ref{qhat-right}) below).

The evaluation of the correction $\delta\hat q$ from  Eq.~(\ref{qhat-rad}) meets with logarithmic divergences. These arise from the region $z\lesssim 1$. 
To the leading-logarithmic accuracy, we can set $z=1$ everywhere, except in the
dominant singularity. Thus the dominant contribution to $\delta\hat q$ can be then written as
\beq
\int_x^1 \rmd z\,\frac{\rmd\delta\hat q (z,xp^+_0, \k^2)}{\rmd 
z}\, D\left(\frac{x}{z},\frac{\k}{z},t_L\right)\,\simeq\,\delta\hat q (x,\k^2)\,
D\left(x,\k,t_L\right)\,,\eeq
with
 \begin{align}\label{qhat-rad2}
\delta\hat q (x,\k^2)\,\equiv \int_{x}^1 \rmd z\,\frac{\rmd\delta\hat q (z,xp^+_0, \k^2)}{\rmd 
z}\,=\, 2\alpha_s \int_{x}^1 \rmd z \int_{\Q,\l}\,
 \big[(\Q+\l)^2 - \l^2\big]\,
{\cal K}\left(\Q,\l,z,xp^+_0\right),
 \end{align}
where the lower limit $x$ in the integral over $z$, which was  a priori present only 
in the `gain' term, has also been inserted in the `loss' term,  while at the same time
multiplying the latter by a factor of 2, to account for its original singularities at both $z=0$ and $z=1$ (which is legitimate since, to the accuracy of interest,
the integral is controlled by values $z\simeq 1 \gg x$). The particular combination of momenta, $\big[(\Q+\l)^2 - \l^2\big]$, that emerges then in Eq.~(\ref{qhat-rad2}) can be given the following interpretation:
when $z\simeq 1$, $\Q \equiv\k - z(\q+\l)$ 
is the same as (minus) the
transverse momentum $\q+\l-\k$ of the unmeasured  daughter gluon. Hence $\Q+\l\simeq \k-\q$ is 
the change in transverse momentum at the emission vertex, with two obvious components:
the momentum $\l$ acquired via medium rescattering during the branching process and the momentum
$\Q$ taken away by the unmeasured daughter gluon. The above applies to the `gain' term.
For the `loss' term, there is no real emission, so the only source of momentum broadening
is the momentum $\l$ transferred from the medium. The difference $(\Q+\l)^2 - \l^2$
represents therefore the net change in the transverse momentum squared, and the average of this quantity over the (momentum dependent) splitting kernel yields the correction $\delta\hat q$.

The complete calculation of the integral (\ref{qhat-rad2}) is presented in Appendix B, where it is shown that the result is dominated by the contribution of the single scattering to the splitting kernel. One gets
 \begin{align}\label{qhat-right}
\delta\hat q (\k^2)=\frac{\alpha_s \,N_c }{2\pi}\, \hat q_0\,
\ln^2\frac{\k^2}{\hat q \tau_{min}}, 
\end{align}
where $\tau_{min}$ is the inverse of the maximum energy that can be extracted from the medium in a single scattering (e.g. $\tau_{min} = 1/T$ for a weakly coupled plasma with temperature $T$). 
This result agrees with that obtained in Ref.~\cite{Liou:2013qya} using a different approach.

The net result of incorporating this large radiative correction is a transport equation
similar to that obtained at leading order, \eqn{Dkt10}, but with an enhanced jet quenching
coefficient, which includes the correction in \eqn{qhat-right}~: 
 \begin{align}\label{Dkt-diff}
\frac{\partial}{\partial t_L}D(x,\k,t_L)&\,=\,
\alpha_s\int_0^1\rmd z\,\bigg[\frac{2}{z^2}\,{\cal K}\left(z,\frac{x}{z}p^+_0\right)
D\left(\frac{x}{z},\frac{\k}{z},t_L\right)-
{\cal K}\left(z,xp^+_0\right)D\left(x,\k,t_L\right)\bigg]\nn
&\quad +\frac{1}{4}\,\left(\frac{\del}{\del\k}\right)^2 \big[\big(\hat q(\k^2)
+\delta\hat q (\k^2)\big)\,D\left(x,\k,t_L\right)\big]\,.
\end{align}
Note that the scale $\k^2$ which controls the size of the double logarithm is the transverse
momentum accumulated by the gluon throughout the medium, that is  $\k^2\sim Q_s^2=\hat q L$.
Hence the argument of the logarithm is large, $\sim L/\tau_{min}$, which makes this radiative corrections
particularly significant.

\subsection{A logarithmic correction to $\hat q$}

The correction that we have exhibited in the previous subsection appears to be the leading correction to the transport coefficient. There are also subleading (logarithmic, instead of double logarithmic) corrections. These have been estimated in Ref.~\cite{Liou:2013qya}, and could in principle be extracted as well from our calculation. In this section, we shall just focus on one particular logarithmic correction that is easy to obtain because it is the correction that naturally emerges when one uses the kernel integrated over $\l$ but not over $\Q$, namely the expression (\ref{Klza}). The starting point is now the equation
\begin{align}\label{Dkt1}
\frac{\partial}{\partial t}D(x,\k,t)&=\int_{\l}\,
{\cal C}(\l,t)D\left(x,\k-\l,t\right)\nn
&+\alpha_s\int_0^1\rmd
z\int_{\q} \bigg[2{\cal K}\left(\Q,z,\frac{x}{z}p^+_0\right)
D\left(\frac{x}{z},\q,t\right)-{\cal K}\left(\q,z,xp^+_0\right)D\left(x,\k,t\right)\bigg]\,,
\end{align}
where $\Q\equiv\k - z\q$.

Expanding the distribution around the momentum $\k$ as in the previous subsection, one gets
\beq\label{eqDkt3b}
&&\frac{\partial}{\partial t}D(x,\k,t)-\frac{1}{4}\,\left(\frac{\del}{\del\k}\right)^2\big[\hat q(\k^2) D\left(x,\k,t\right)\big]\nn
&&=\alpha_s\int_0^1\rmd
z \,\Bigg[\frac{2}{z^2}{\cal K}\left(z,\frac{x}{z}p^+_0\right)
D\left(\frac{x}{z},\frac{\k}{z},t\right) -{\cal K}\left(z,xp^+_0\right)D\left(x,\k,t\right)\bigg]
\nn
&&+ \alpha_s\int_0^1\rmd
z\int_\Q \frac{2}{z^2}{\cal K}\left(\Q,z,\frac{x}{z}p^+_0\right)
\frac{1}{4} Q^2\frac{\del^2}{\del \k^2}D(x,\k)
\eeq
 The second--order term in the expansion (which carries
the divergence near $z=1$) yields a correction  to $\hat q$, which we call $\delta\hat q'$. We get
(below, we use approximations valid for $z\simeq 1$)
 \beq\label{qhat-mult}
\delta\hat q'&=& 2\alpha_s \int_{x}^1 \rmd z \int_{\Q}\,
 \Q^2\,
{\cal K}\left(\Q,z,xp^+_0\right)\nn
 &\propto&\alpha_s \int_{x}^1 \rmd z \,k_{_{\rm br}}^4(z,xp^+_0)\,
 \frac{P(z)}{(1-z)xp_0^+}\,\sim\, \alpha_s \hat q \int_{x}^1 \frac{\rmd z}{1-z}\,,
  \eeq
where to obtain the estimate in the second line we have used the fact that the splitting
kernel is peaked at
$k_{_{\rm br}}^2(z,xp^+_0)= \sqrt{(1-z)xp_0^+\hat q}$, cf.  \eqn{Klza}. 
As anticipated, there is a logarithmic divergence
at $z=1$, corresponding to $\omega_0\to 0$. This must  be cut at the lowest energy scale at which
the BDMPSZ mechanism is applicable, which  is the Bethe--Heitler energy $\omega_{_{\rm BH}}
\equiv \hat q_0 \lambda_{_{\rm mfp}}$, i.e. the energy for which the branching time 
$\tau_{_{\rm br}}(\omega_0)$ becomes of the order of the mean free path $\lambda_{_{\rm mfp}}$. 
In practice this means that the integral over $z$  in \eqn{qhat-mult}
 must be restricted to $1-z\le
\omega_{_{\rm BH}}/\omega$, with $\omega=xp_0^+$ the energy of the measured gluon. Note that single scattering 
does not contribute to this logarithmic correction (as it can be checked using \eqn{kernel5b}), in contrast to the double logarithmic one discussed in the previous subsection.  Let us also emphasize that \eqn{qhat-mult} is only one
among the several logarithmic corrections to $\hat q$ that have been analyzed in Ref.~\cite{Liou:2013qya}.

\section{Conclusions}

In this paper, we have extended our previous studies of the in-medium QCD cascade, based on the approximation that successive gluon branchings can be treated as independent from each other. This approximation is 
indeed justified for the typical partons within the cascade, whose formation times are much smaller than the medium 
size. We have constructed a generating functional for the various relevant probabilities and deduced from it the evolution equation for the  inclusive one-gluon distribution function, that keeps track of the transverse momentum of the measured gluon. In this equation, however, the transverse momenta entering the splitting kernel are treated in an average way and the splittings are effectively collinear.  This is justified since the transverse 
momentum broadening during the comparatively short (in our approximation, quasi--instantaneous) 
branching processes is much smaller than that accumulated via collisions in the medium 
all the way along the parton trajectories. By relaxing some of our approximations, in particular those which allow us to integrate the kernel over transverse momenta, we were able to identify large corrections to the jet quenching parameter, and in particular to recover the double logarithmic contribution that has been calculated recently in a general study of transverse momentum broadening.

\section*{Acknowledgments}

We would like to thank Al Mueller and Bin Wu for many useful discussions and comments on the
manuscript.
This research is supported by the European Research Council under the Advanced Investigator Grant ERC-AD-267258.

\appendix

\section{The splitting kernel}
\begin{figure}
\begin{center}
\includegraphics[height=10cm]{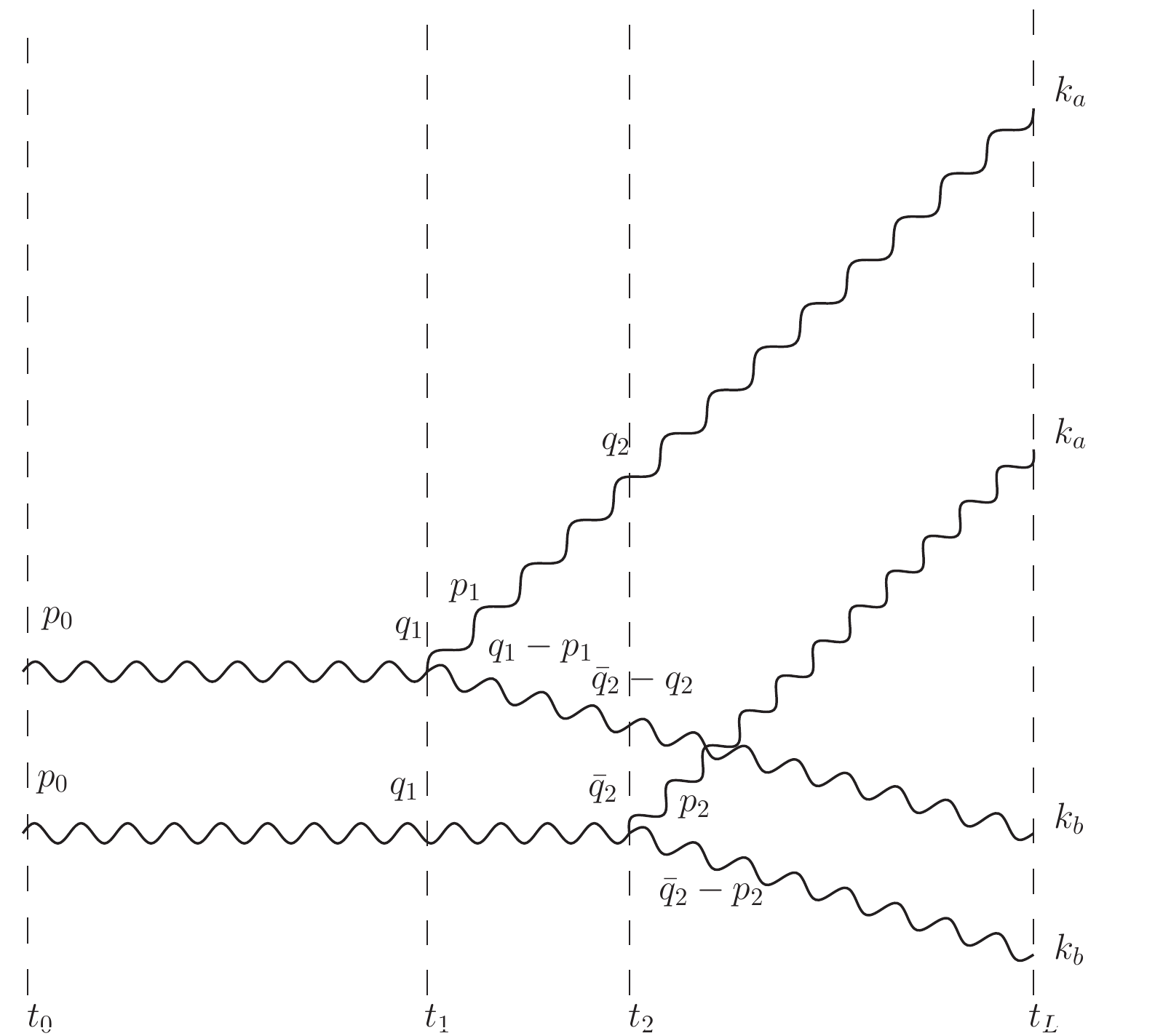}
\caption{The momenta of the intermediate states in Eq.~(\ref{Sigma2b}). The amplitude is drawn above the complex conjugate amplitude (see Ref.~\cite{Blaizot:2012fh}), with the gluon splitting occurring at time $t_1$ in the amplitude, and at time $t_2$ in the complex conjugate amplitude. At any given time, the sum of momenta in the amplitude equals that of momenta in the complex conjugate amplitude.} \label{figure3}
\end{center}
\end{figure}

 It was shown  in Ref.~\cite{Blaizot:2012fh} that the cross section for observing at time $t_L$  two gluons with momenta ${\vec k_a}$, $\vec{k}_b$, given that a single gluon was present with momentum $\vec{p}_0$ at time $t_0$, is given, in leading order perturbation theory, by 
\beq
\frac{\rmd^2\sigma}{\rmd\Omega_{k_a}\rmd\Omega_{k_b}}= \int{\rm d}\Omega_{p_0} \,P_2(\vec{k}_a,\vec{k}_b;t_L,t_0)\, \frac{d\sigma_{hard}}{d\Omega_{p_0}},
\eeq
where ${d\sigma_{hard}}/{d\Omega_{p_0}}$ is the hard cross section for the production of the initial gluon, and $\rmd\Omega_k\equiv (2\pi)^{-3} \rmd^2\k\, 
\rmd k^+/2k^+$ is the invariant  phase--space element, and the vector notation $\vec{k}$ stands for $(p^+,\p)$, as in the main text of this paper. The probability $P_2(\vec{k}_a,\vec{k}_b;t_L,t_0)$ can be written as (cf. Eq.~(\ref{defP2}))
\beq
P_2(\vec{k}_a,\vec{k}_b;t_L,t_0)=2\pi\,2p_0^+\delta(k_a^++k_b^+-p_0^+)\,{\cal P}_2(\k_a,\k_b,z;t_L,t_0),
\eeq
with $z=k_a^+/p_0^+$,  and ${\cal P}_2(\k_a,\k_b,z;t_L,t_0)$  given by 
\beq\label{Sigma2b}
&&{\cal P}_2(\k_a,\k_b,z;t_L,t_0)=
\frac{g^2 P_{gg}(z)}{z(1-z)(p^+_0)^2}\,2\Re e\,\int_{t_0}^{t_L}\rmd t_2\int_{t_0}^{t_2}\rmd t_1\,\int_{ \p_1\q_1\bar\q_2 \p_2\q_2} \,(\hat\P_1\cdot\hat\Q_2)\,\nn
&&\quad \times\,(\k_a\k_b;\k_a\k_b|\tilde S^{(4)} (t_L,t_2)|\q_2,\bar\q_2-\q_2;\p_2,\bar\q_2-\p_2)(\q_2,\bar\q_2-\q_2;\bar\q_2| \tilde S^{(3)}(t_2,t_1)|\p_1,\q_1-\p_1;\q_1)\nn
&&\quad\times(\q_1;\q_1|\tilde S^{(2)}(t_1,t_0)|\p_0;\p_0),
\eeq
where $\hat \P_1\equiv \p_1-z\q_1$, $\hat \Q_2\equiv \p_2-z\bar\q_2$ denote the  `natural' momentum variables\footnote{The momentum $\hat \P_1=\p_1-z\q_1$ is the relative momentum of the non relativistic motion of the two offspring gluons in the transverse plane. Alternatively, $|\hat\P_1|/p_1^+\simeq \theta_{z}$ is the polar angle of the gluon carrying $zq_1^+$. Since $|\hat\P_1|\simeq k_{\rm br}$, $\theta_z\sim k_{\rm br}/z q_1^+$.} at the vertices in the amplitude and the complex conjugate amplitude, respectively. The explicit flow of momenta that label intermediate states is illustrated in Fig.~\ref{figure3}. The dependence of ${\cal P}_2$ (and $P_2$) on the initial momentum $\vec{p}_0$ is left implicit to simplify the notation. 
The real part takes into account the  time ordering $t_1>t_2$ not explicitly included in (\ref{Sigma2b}). 
The formula above has been obtained after performing the average over the field fluctuations using Eq.~(\ref{correlmed}),  and summing over polarizations. Azymuthal angles of the  momenta at the vertices have also been averaged. 

At this point no approximation has been made, except for  the obvious restriction to leading order in perturbation theory (in the background field), that is, a single splitting occurs between $t_0$ and $t_L$.
One may now introduce several approximations that are valid in the regime where the branching occurs on a time scale that is small compared to the length of the medium, i.e, in the regime $\tau_{_{\rm br}}\ll t_L-t_0$ (or equivalently for infinite medium length). We shall first consider the following two approximations:

1)  Ignore the non factorizable piece of $\tilde S^{(4)}$, that is, set
\beq
&&(\k_a\k_b;\k_a\k_b|\tilde S^{(4)}_{\rm fac} (t_L,t_2)|\q_2,\bar\q_2-\q_2;\p_2,\bar\q_2-\p_2)\nn
&&\qquad=(2\pi)^2\delta^{(2)}(\p_2-\q_2){\cal P}(\k_a-\q_2,t_L,t_2){\cal P}(\k_b-\bar\q_2+\q_2,t_L,t_2).\label{S4tildefac}
\eeq
It was shown in  \cite{Blaizot:2012fh} that the non factorizable piece of the 4-point function  dies away over a time scale of order $\tau_{_{\rm br}}$, and it is  down by a least one power of $\tau_{_{\rm br}}/L$ as compared to the factorized part. 

2) Use as time integration variables $t_1$ and $t_2-t_1\equiv\tau$, i.e., set $t_2=t_1+\tau$, and neglect $\tau$ in the ${\cal P}$ factors that enter the 4-point function (\ref{S4tildefac}), that is e.g.
\beq
{\cal P}(\k_a-\q_2,t_L,t_1-\tau)\rightarrow {\cal P}(\k_a-\q_2,t_L,t_1),
\eeq
and similarly for the other ${\cal P}$. This allows us to integrate freely the 3-point function over $\tau$ from 0 to $\infty$ (as we shall soon recall,  the 3-point function is strongly damped as soon as $\tau\gtrsim \tau_{_{\rm br}}$).
\\
With these two approximations, Eq.~(\ref{Sigma2b}) simplifies to
\beq\label{Sigma2fac-2a}
&&{\cal P}_2(\k_a,\k_b,z;t_L,t_0)=2g^2z(1-z) 
\,\int_{t_0}^{t_L}\rmd t_1\,\int_{\q_1,\hat\Q_2,\l} \;{\cal P}(\k_a-\q_2,t_L,t_1){\cal P}(\k_b-\q_1-\l+\q_2,t_L,t_1)\nn
& &\quad\quad\times \frac{P_{gg}(z)}{\omega_0^2}\, \Re e\left[   \int_{\hat\P_1}\int_0^\infty \rmd \tau \,(\hat\P_1\cdot\hat\Q_2)\,(\q_2,\bar\q_2-\q_2;\bar\q_2| \tilde S^{(3)}(t_2,t_1)|\p_1,\q_1-\p_1;\q_1) \right]\nn
 & &\quad\quad\times  \, \,{\cal P}(\q_1-\p_0,t_1-t_0),
\eeq
where we have set $\omega_0\equiv z(1-z)p_0^+$ and
we have used as independent variables $\q_1, \hat \P_1=\p_1-z\q_1, \hat\Q_2=\p_2-z\bar\q_2=\q_2-z\bar\q_2, \l=\bar\q_2-\q_1$ in place of $\q_1, \p_1, \bar\q_2,\q_2$. 
At this point we set (with a slight abuse of notation)
\beq
\tilde S^{(3)}(\hat\P_1,\hat\Q_2, \l,z,\tau,t_1)=(\q_2,\bar\q_2-\q_2;\bar\q_2| \tilde S^{(3)}(t_2,t_1)|\p_1,\q_1-\p_1;\q_1)
\eeq
which makes explicit the relevant momentum variables on which the 3-point function depends, and we define the splitting kernel
 \beq\label{kernel5}
{\cal K}(\hat \Q,\l,z,t_1)\,\equiv\, \frac{P_{gg}(z)}{\omega_0^2}\, {\rm Re}
\int_{0}^{\infty} \rmd\tau \int_{\hat\P} (\hat\P\cdot\hat\Q)\,  \tilde S^{(3)}(\hat\P,\hat\Q, \l,z,\tau,t_1).
\eeq 
With this new notation, Eq.~(\ref{Sigma2fac-2a}) reads
\beq\label{Sigma2fac-2b}
&&{\cal P}_2(\k_a,\k_b,z;t_L,t_0)=2g^2z(1-z) 
\,\int_{t_0}^{t_L}\rmd t_1\,\int_{\q_1,\hat\Q_2,\l} \;{\cal P}(\k_a-\hat\Q_2-z(\q_1+\l),t_L,t_1)\nn
& &\quad\quad\times {\cal P}(\k_b+\hat\Q_2-(1-z)(\q_1+\l),t_L,t_1)\;{\cal K}(\hat \Q_2,\l,z,t_1)
  \, {\cal P}(\q_1-\p_0,t_1,t_0).
\eeq

At this point further approximations are legitimate. For instance, as we did in \cite{Blaizot:2012fh}, we can neglect the momentum $\l$ in the ${\cal P}$ factors: indeed $\l$ represents the typical momentum acquired during the branching process, $\l^2\simeq \hat q \tau_{_{\rm br}}$, and it is small compared to $k_a$ or $k_b$ which are both of order $Q_s\sim \hat q(t_L-t_0)$. If one neglects $\l$ in the ${\cal P}$ factors, then one can integrate the splitting kernel over $\l$ and get the simpler formula
\beq\label{Sigma2fac-2b2}
&&{\cal P}_2(\k_a,\k_b,z;t_L,t_0)=2g^2z(1-z) 
\,\int_{t_0}^{t_L}\rmd t_1\,\int_{\q_1,\hat\Q_2} \;{\cal P}(\k_a-\hat\Q_2-z\q_1,t_L,t_1)\nn
& &\quad\quad\times {\cal P}(\k_b+\hat\Q_2-(1-z)\q_1,t_L,t_1)\;{\cal K}(\hat \Q_2,z)
  \, {\cal P}(\q_1-\p_0,t_1,t_0).
\eeq
with ${\cal K}(\hat \Q,z,t)\equiv\int_\l {\cal K}(\hat \Q,\l,z,t)$. This is the approximation that was explicitly considered in Ref.~\cite{Blaizot:2012fh}.

We may also observe that the variable $\hat \Q$ that stands as argument of ${\cal K}$ is also small, of order $k_{_{\rm br}}\ll Q_s$, and can also be neglected in a leading order approximation. Doing so, one ends up with an even simpler formula 
\beq\label{Sigma2fac-2b3}
&&{\cal P}_2(\k_a,\k_b,z;t_L,t_0)=2g^2z(1-z) 
\,\int_{t_0}^{t_L}\rmd t_1\,{\cal K}(z,t_1)\int_{\q_1} \;{\cal P}(\k_a-z\q_1,t_L,t_1)\nn
& &\quad\quad\times {\cal P}(\k_b-(1-z)\q_1,t_L,t_1)\;
  \, {\cal P}(\q_1-\p_0,t_1,t_0).
\eeq
with ${\cal K}(z,t)\equiv\int_{\hat\Q}{\cal K}(\hat \Q,z,t)=\int_{\Q \l}{\cal K}(\hat \Q,\l,z,t)$ is the fully integrated splitting kernel. This is the kernel used to construct the generating functional of the in-medium cascade in Sect.~3.
\\

As was shown in \cite{Blaizot:2012fh}, the 3-point function can be written as the following path integral
\beq\label{S3}
&&\tilde S^{(3)}(\P,\Q, \l,z,p_0^+;t_2,t_1)=\int \rmd\u_1 \rmd\u_2 \rmd\v \; \rme^{i\u_1\cdot\P-i\u_2\cdot\Q-i\v\cdot\l}\nn
&&\times\int_{\u_1}^{\u_2} {\cal D}\u\exp\left\{\frac{i\omega_0}{2} \int_{t_1}^{t_2} \rmd t ~\dot{\u}^2 -\frac{N_c}{4}\int_{t_1}^{t_2} \rmd t\,  n(t) \left[\sigma(\u)+\sigma(\v-z\u)+\!\sigma(\v+(1-z)\u)\right]\right\}.\nn
\eeq
This can be explicitly evaluated within the `harmonic 
approximation', which assumes $\sigma(\r)\propto \hat q \r^2$ (cf. \eqn{sigma}). By expanding all the $\sigma$'s to quadratic order, and performing the resulting gaussian path integrals, one gets \cite{Blaizot:2012fh}\footnote{Note that a mistake was made in evaluating the Gaussian path-integral in \cite{Blaizot:2012fh}: in going from Eq.~(B.24)
 to Eq.~(B.25) in Appendix B of Ref.~\cite{Blaizot:2012fh}, one has ignored the shift in the endpoints of the trajectory $\u(t)$. This was of no consequence in \cite{Blaizot:2012fh} since the splitting kernel was there integrated over $\l$. However this affects the $\l$ dependence of the kernel, which is here given correctly. }
\beq\label{S3harm}
&& \tilde S^{(3)}(\P,\Q, \l,z,p_0^+;t_2,t_1)=\frac{16\pi f(z)}{3\hat q\Delta t}\exp\left\{-\frac{4f(z)
\big[\l+(1-2z)(\P-\Q)/2f(z)\big]^2}{3\hat q\Delta t}\right\}\nn
&&\quad\times\ \frac{2\pi i}{\Omega\omega_0\sinh(\Omega\Delta t)}\exp\left\{-i\frac{(\P+\Q)^2}{4\omega_0\Omega\coth(\Omega\Delta t/2)}-i\frac{(\P-\Q)^2}{4\omega_0\Omega\tanh(\Omega\Delta t/2)}\right\}\,.
\eeq
where $\Delta t\equiv t_2-t_1$, $f(z)\equiv 1-z(1-z)$, and
\beq
\label{Omega}
\Omega\equiv\, \frac{1+i}{2\tau_{_{\rm br}}(z,p^+_0)},\eeq
with
$\tau_{_{\rm br}}(z,p^+_0)
\equiv \sqrt{{\omega_0}/{\hat q_0}}$ and $\hat q_0\equiv \hat q f(z)$.

By inserting the result \eqref{S3harm} for $\tilde S^{(3)}$ into \eqn{kernel5}, one finds the following
integral representation for the splitting kernel (in the harmonic approximation):
\begin{align}
\label{kernel-harmonic}
&{\cal K}(\Q,\l,z,p_0^+;t)\,= \, {16\pi}\,
\frac{f(z) P_{gg}(z)}{\omega_0^2} \, \Re e \int_0^\infty \frac{\rmd\Delta t}
{3\hat q\Delta t}\, \nn
&\times \int_{\P} \,( \Q\cdot\P)\,{\exp}\left\{ -\frac{4f(z)\,
\big[\l+(1-2z)(\P-\Q)/2f(z)\big]^2}{3\hat q\Delta t} \right\} \nn
&\times \frac{2\pi i}{\Omega \omega_0 \sinh(\Omega \Delta t) }
\exp{\left\{  -i\frac{(\P+\Q)^2}{4\omega_0\Omega\coth(\Omega \Delta t/2)}
-i\frac{(\P-\Q)^2}{4\omega_0\Omega\tanh(\Omega \Delta t/2)}\right\}  }.
\end{align}
This kernel obeys the symmetry property:
 \beq\label{symK}
 {\cal K}(\Q,\l,z,p_0^+;t)\,=\,{\cal K}(-\Q,\l,1-z,p_0^+;t)\,,\eeq
which expresses the symmetry of the splitting under the exchange of the two daughter gluons.

By performing the integration over $\l$ one recovers the splitting kernel obtained in  \cite{Blaizot:2012fh}:
\beq\label{Klza}
{\cal K}(\Q,z,p_0^+;t)\equiv\int_\l {\cal K}(\Q,\l,z,p_0^+; t)=\frac{2}{p_0^+}\frac{P_{gg}(z)}{z(1-z)}\,
\sin\left[\frac{\Q^2}{2k_{_{\rm br}}^2}\right]\exp\left[-\frac{\Q^2}{2k_{_{\rm br}}^2}\right],
\eeq
In this expression,
 $k_{_{\rm br}}^2=\hat q_0 \tau_{_{\rm br}}(z,p^+_0)=\sqrt{\omega_0\hat q_0}$
is the typical transverse momentum squared transferred via medium rescattering during the splitting. Note that the branching time, and hence the splitting kernel, depend upon both $p_0^+$ 
(the energy of the parent gluon) and upon the splitting fraction $z$.
The expression (\ref{Klza})  illustrates an important property of the branching that is induced by soft multiple collisions: it is strongly peaked at  $|\Q|\sim k_{_{\rm br}}(z,p^+_0)$. 
For smaller momenta $|\Q|\ll k_{_{\rm br}}$, gluon splitting  is suppressed by  $Q^2/k_{_{\rm br}}^2$, which reflects the interferences of the LPM effect. At  larger momenta $|\Q|\gtrsim k_{_{\rm br}}$,
it is rapidly damped, as  it is unlikely to acquire more transverse momentum than $k_{_{\rm br}}$ via multiple scattering. 

The kernel in \eqn{Klza} contains information about the geometry of the medium--induced splitting: 
 the polar angles made by the two
offspring gluons with respect to their parent parton are $\theta_z\simeq |\Q|/zp_0^+$
and respectively $\theta_{1-z}\simeq |\Q|/(1-z)p_0^+$.  Since $|\Q|\sim k_{_{\rm br}}\ll Q_s=\sqrt{\hat q L}$,
it is clear that these angles are negligible compared to the angular spreading acquired via collisions
in between successive branchings. By integrating the kernel (\ref{Klza}) over $\Q$, this information about
the emission angles is averaged out, and 
one obtains the fully integrated kernel ${\cal K}(z,p_0^+;t)$ given explicitly in Eq.~(\ref{Kz2}).

Finally, we shall write the expression of the splitting kernel in the limit where a single scattering occurs during the branching process. We limit ourselves to the case where $z\lesssim1$,  the case of relevance for discussing the double logarithmic correction to $\hat q$. The three-point function in the one-scattering approximation, obtained by expanding (\ref{S3}) to leading order in $\sigma$, takes the form
\beq
&&\tilde S^{(3)}(\P,\Q, \l)\approx -\frac{N_c}{4}\int_{t_1}^{t_2} {\rmd}t\,n(t)\, {\cal G}_0(\Q,t_2-t)\,{\cal G}_0(\P,t-t_1)
\nn
&&\times \left[ (2\pi)^2\delta(\l)\,\sigma(\Q-\P)+(2\pi)^2\delta(\Q-\P+\l)\,\sigma(\l)
+(2\pi)^2\delta(\Q-\P)\,\sigma(\l)\right],\nn 
\eeq
where ${\cal G}_0(\Q,t_2-t_1)$ is the free propagator
\beq
{\cal G}_0(\Q,t_2-t_1)= {\rm e}^{-i\frac{\Q^2}{2\omega_0}(t_2-t_1)}.
\eeq
When needed (see below) a small negative imaginary part may be added to $\Q^2$ to account for the retarded condition. If we assume that $n$ is independent of time, we can perform the time integration, and obtain ($\tau\equiv t_2-t_1$)
\beq\label{3pmom}
&&\tilde S^{(3)}(\P,\Q, \l)\approx - \frac{N_c n\omega_0}{2i}\,\frac{   \rme^{-i\frac{\P^2}{2\omega_0}\tau}- \rme^{-i\frac{\Q^2}{2\omega_0}\tau} }{ Q^2-P^2  }
\nn
&&\times \left[ (2\pi)^2\delta(\l)\,\sigma(\Q-\P)+(2\pi)^2\delta(\Q-\P+\l)\,\sigma(\l)
+(2\pi)^2\delta(\Q-\P)\,\sigma(\l)\right].\nn 
\eeq
At this point,  the kernel reads
\beq\label{kernel5b}
&&{\cal K}(\Q,\l,z)\,\equiv\, \frac{ C_A^2 n}{1-z}\, {\rm Re} 
\int_{\P} \frac{\P\cdot\Q}{Q^2P^2}\nn
&&\times \left[ (2\pi)^2\delta(\l)\,\sigma(\Q-\P)+(2\pi)^2\delta(\Q-\P+\l)\,\sigma(\l)
+(2\pi)^2\delta(\Q-\P)\,\sigma(\l)\right].\nn 
\eeq 
where we have used the time integral (recall that $P^2\to P^2-i\epsilon$)
\beq
\int_0^\infty \rmd \tau \, \frac{   \rme^{-i\frac{\P^2}{2\omega_0}\tau}- \rme^{-i\frac{\Q^2}{2\omega_0}\tau} }{ Q^2-P^2  }=-\frac{2i\omega_0}{Q^2 P^2}.
\eeq
This expression will be used in the next Appendix. Note that the last two terms in the r.h.s of Eq.~(\ref{kernel5b}) vanish upon integration over $\l$, because of the identities (\ref{sigmaproperties}). These terms play an essential role in the evaluation of $\delta \hat q$, as shown in the next Appendix.

\section{Estimating the double logarithmic correction to $\hat q$}

Our starting point is the integral representation for the kernel given, in the harmonic approximation, by Eq. (\ref{kernel-harmonic}),
where we keep only the singular part at $z\to1$  from the Altarelli--Parisi splitting function $P_{gg}(z)$ (see Eq.~(\ref{Pgg})), and 
we set $z=1$ in the rest of the expression. 
After inserting Eq. (\ref{kernel-harmonic}) into the r.h.s. of \eqn{qhat-rad2},
we are facing four integrations: an integration over the duration $\tau$ of the branching process 
and three Gaussian integrations over the momentum variables $\l$, $\Q$ and $\P$. Performing first 
the integral over $\l$, one obtains (with $\tau\equiv \Delta t$)
\beq
&&\delta\hat q(\k^2)\simeq
2\alpha_s \, \Re e 
\int_{x}^1\frac{dz}{1-z} \frac{C_A}{\omega^2_0} \int_0^\infty \rmd \tau  \int_{\Q,\P}
\, (\Q\cdot\P)^2\,\nn
&&\times \frac{2\pi i}{\omega_0\Omega\sinh(\Omega \tau) }\exp{\left\{  -i\frac{(\P+\Q)^2}{4\omega_0\Omega\coth(\Omega  \tau/2)}-i\frac{(\P-\Q)^2}{4\omega_0\Omega\tanh(\Omega  \tau/2)}\right\}  },
\eeq
where
$\omega_0\equiv (1-z)zq^+ \simeq (1-z)xp^+_0$ and $\omega_0\Omega^2=i\hat q_0/2$ (cf. \eqn{Omega}). Note
that for $z$ close to one, the quantity $\omega_0$ is essentially the energy $(1-z)q^+$ of the unresolved gluon and $\hat q_0\simeq\hat q$.
It is now straightforward to perform the remaining momentum integrations, yielding
\beq\label{q11}
&&\delta\hat q(\k^2)=2 \, \Re e 
\int_{x}^1\frac{dz}{1-z} \frac{\alpha_s \,C_A}{\omega^2_0} \int_0^\infty \rmd \tau  \,\frac{i (\omega_0\Omega)^3}{\pi}\frac{1}{\sinh(\Omega \tau)}\left[1+\frac{4}{\sinh^2(\Omega  \tau)}\right].
\eeq
Anticipating on the fact that the dominant (divergent) contribution will come from the small $\tau$ region, we carefully expand the integrand  for $|\Omega|\tau\ll 1$ and get 
 \beq\label{smalltau}
\frac{ \Omega^3}{\sinh \Omega\tau}\,\left(1+\frac{4}{\sinh^2 \Omega\tau} \right)\approx \frac{4}{\tau^3}-\frac{\Omega^2}{\tau}.
\eeq
The $\Omega$--independent piece is real, so it does not contribute to the real part
of the integral in \eqn{q11} (because of the explicit factor $i$ in Eq.~(\ref{q11})). The second term in the r.h.s. \eqn{smalltau} is purely
imaginary and is linear in $\hat q_0$, suggesting that it describes the contribution of a {\em single scattering} (see below). 
When inserted into \eqn{q11}, this term generates an integral which is logarithmically divergent
as $\tau\to 0$. One then gets, after changing the
integration variable from $z$ to $\omega_0= (1-z)xp^+_0$,
\beq\label{qhat-DL0} 
\delta\hat q (\k^2)&\approx  &\frac{\alpha_s \,N_c }{\pi}\, \hat q\,
\int  \frac{\rmd\omega_0}{\omega_0}\int 
\frac{\rmd\tau}{\tau}\,.
\eeq

We can verify that the dominant contribution to ${\cal K} $ is coming from a single scattering with the medium by using the expression obtained in Appendix A, Eq.~(\ref{kernel5b}),  in order to perform the calculation. We get  (with $\omega_0=(1-z)p^+$)
\beq\label{correctionqhat1}
\omega_0\frac{\rm d\delta\hat q }{\rmd \omega_0}&=&2\alpha_s \int_{\Q,\l}\, \q\cdot(\Q+2\l)\,{\cal K}(\Q,\l,z)\nn
&=& 4 \alpha_s N_c^2 n\,{\rm Re} \int_{\Q,\P}\frac{(\Q\cdot\P)^2}{\Q^2\P^2}\sigma(\Q-\P)
\eeq
where, in order to  perform the $\l$ integration we have used the identities (\ref{sigmaproperties}).
By using Eq.~(\ref{FTsigmadip}), we get
\beq
\int_{\Q,\Q}\frac{(\Q\cdot\P)^2}{\Q^2\P^2}\sigma(\Q-\P)&=&2g^2\int_{\q,\l}\,\gamma(\l)\frac{\Q^2\l^2-(\l\cdot\Q)^2}{\Q^2(\Q-\l)^2}\nn
&\approx &g^2 \int_{\Q}\,\frac{1}{\Q^2}\, \int_\l \l^2 \gamma(\l),
\eeq
where, in the last line,  we have made approximations valid in the region $l\ll Q$. We therefore get from (\ref{correctionqhat1})
\beq
\omega_0\frac{\rm d\delta\hat q}{\rmd \omega_0}
&=& 2 \alpha_s N_c^2 n\, \int_{\Q}\,\frac{1}{\Q^2}\, \int_\l \l^2 V(\l)\nn
&=&\frac{\alpha N_c}{\pi}\,\hat q\,\int\frac{\rmd \Q^2}{\Q^2},
\eeq
which is recognized as Eq.~(\ref{qhat-DL0}) after recalling that the
formation time and the virtuality of the soft emitted gluon are related by
$\tau\simeq \omega_0/Q^2$.

Returning to Eq.~(\ref{qhat-DL0}), we shall now discuss
the boundaries of the double integral there. Consider the time integral first. 
At larger times $|\Omega|\tau\gtrsim 1$, this integral is cutoff by the
exponential decay of $[\sinh \Omega\tau]^{-1}$~; this is the effect of {\em multiple scattering} during the
emission process, which limits
the branching times to values $\tau \lesssim \tau_{_{\rm br}}(\omega_0)=\sqrt{\omega_0/\hat q}$. On the other hand the time $\tau$ cannot be smaller that the inverse of the maximum energy that can be taken away from the medium through a single scattering \cite{Liou:2013qya}. We call this limiting time
$\tau_{min}$. This lower bound may not always  be reached however. If $\omega_0$ is not too small, then $\tau$ will be limited by the formation time of the unobserved gluon, $\tau\simeq \omega_0/Q_\perp^2$. But in reality such
values cannot exceed the transverse momentum $k_\perp$ of the measured gluon, which in turn implies
a lower limit $\sim \omega_0/k_\perp^2$ in the integral over $\tau$.
Thus the lower bound on $\tau$ is  ${\rm max}\,\big(\tau_{min},\, \omega_0/k_\perp^2\big)$.

Turning now to $\omega_0$, we note that the lower limit at $z=x$ in the original integral
over $z$ implies an upper limit $\omega_0^{\rm max}=(1-x)xp^+_0\simeq xp^+_0 = \omega$
(the energy of the measured gluon) in the integral over $\omega_0$. For the ensuing integral to have a non--trivial support when $\tau=\tau_{min}$, one also needs
$\tau_{_{\rm br}}(\omega_0)\equiv \sqrt{\omega_0/\hat q}\,\gtrsim\tau_{min}$, that is,
$\omega_0\gtrsim  \hat q \tau_{min}^2$. 

In view of the above, we need to split the integral over $\omega_0$ into two regions:
\beq\label{qhat-DL} 
\hat q_{1} (\k^2)&\approx  &\frac{\alpha_s \,N_c }{\pi}\, \hat q\,\left\{    
\int_{\hat q \tau_{min}^2}^{\tau_{min} \k^2} \frac{\rmd\omega_0}{\omega_0}\int_{\tau_{min}}^{\sqrt{\omega_0/\hat q}} 
\frac{\rmd\tau}{\tau}\,+\int_{ \tau_{min} \k^2}^{\omega} \frac{\rmd\omega_0}{\omega_0}\int_{\omega_0/\k^2}^{\sqrt{\omega_0/\hat q}} 
\frac{\rmd\tau}{\tau}\right\}.\\
&\approx& \frac{\alpha_s \,N_c }{2\pi}\, \hat q\,\left\{ \ln^2\frac{\k^2}{\hat q\tau_{min}}-\frac{1}{2}\ln^2\frac{\k^4}{\hat q\omega}\right\}.
\eeq
Dropping the last term (negligible if $\omega\gg \tau_{min}\k^2$), one  finds the result (\ref{qhat-right}).

\section{Equivalence between forward and backward evolutions}
As mentioned earlier, one may write two types of evolution equations, depending on whether one differentiates the generating functional with respect to the initial of the final times. These two evolutions are referred to as backward and forward Kolmogorov evolutions (see Ref.~(\cite{Cvitanovic:1980ru}) for a general discussion). In section 3, we have discussed the forward case. We discuss here the backward case that is often preferred for Monte-Carlo implementations. The two formulations are in principle equivalent, although the forms of the resulting equations may look rather different. At the end of this Appendix, we shall prove the equivalence in the case of the inclusive one gluon distribution.

In order to derive the backward evolution equation for the generating functional, which we denote now ${\cal Z}[p^+,\p;t_L,t_0|u]$\footnote{In contrast to what happens in the forward evolution, in the present case the variable $p^+$ changes in the evolution.}, we first note the analog of Eq.~(\ref{diffP2L}) for the time derivative of ${\cal P}_2$, with now the derivative acting on $t_0$:
 \beq\label{diffP20}
&&-\del_{t_0}{\cal P}_2(\k_a,\k_b,z;t_L,t_0)\nn
&&\qquad=2g^2z(1-z) 
{\cal K}(z,p_0^+; t_0){\cal P}(\k_a-z\p_0;t_L,t_0){\cal P}(\k_b-(1-z)\p_0;t_L,t_0).
  \eeq
This provides the essential ingredient for the construction of the evolution equation, which reads:
 \beq\label{GFdiff}
-\frac{\partial}{\partial t_0}{\cal Z}[\vec{p};t_L,t_0|u]&=&\alpha_s\int_0^1
\rmd z\,{\cal K}(z,p^+;t_0)\Big\{{\cal Z}[z\vec{p};t_L,t_0|u]\,{\cal Z}[(1-z)\vec{p};t_L,t_0|u]-{\cal Z}[\vec{p};t_L,t_0|u]\Big\}\nn
&+&\int_{\l} {\cal C}(\l,t_0){\cal Z}[p^+,\p-\l;t_L,t_0|u]\,.
\eeq
The term  quadratic in ${\cal Z}$ within the braces in the r.h.s. describes
the splitting of the initial parton into two partons, whereas the term linear in ${\cal Z}$ is
necessary to ensure probability conservation. 
As usual, the collision   term  which involves ${\cal C}$ accounts for 
transverse momentum broadening. 
 
\begin{figure}
\begin{center}
\includegraphics[height=5.2cm]{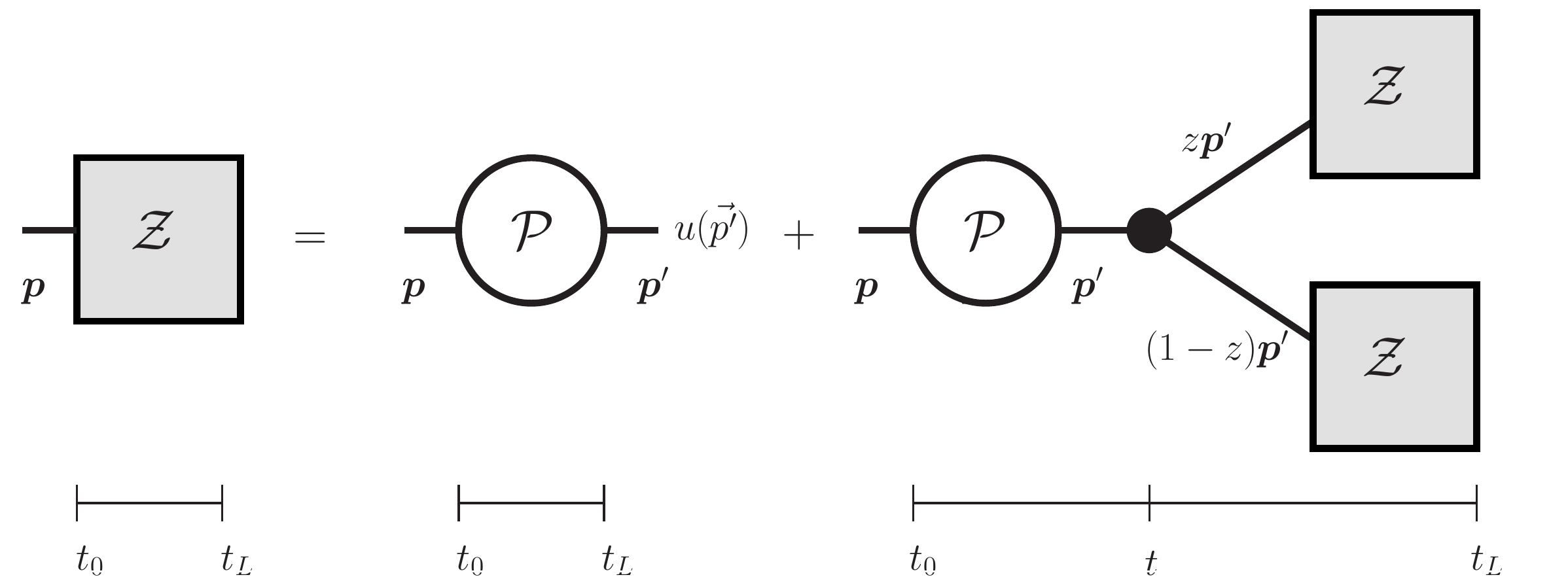}
\caption{Diagrammatic representation of the master equation (\ref{GF})} \label{figure2}
\end{center}
\end{figure}

This differential equation can easily be transformed into an integral equation, which reads
 \beq\label{GF}
&&{\cal Z}[\vec{p}; t_L,t_0|u]=\int_{\p'} \,\Delta(p^+; t_L-t_0)
{\cal P}(\p'-\p; t_L,t_0)\, u(p^+,\p')\nn 
&&+\alpha_s\int_{t_0}^{t_L}\rmd t\,\int_{\p'}\Delta(p^+; t-t_0)
{\cal P}(\p'-\p; t_L,t_0)\nn
&&\times \int_0^1\rmd z\,
{\cal K}(z,p^+,t)\, {\cal Z}\big[zp^+,z\p';t_L,t|u\big]\
{\cal Z}\big[(1-z)p^+,(1-z)\p';t_L,t|u\big]\,,\nn
\eeq
where $\Delta(p^+; t_L-t_0)$ is the Sudakov factor defined in \eqn{Delta}.
This equation, which is graphically illustrated in Fig.~\ref{figure2},
recursively generates the ensemble of the cascade by `inserting one additional splitting at the beginning
of the cascade'. 

As an illustration of the equivalence between the two different versions for the evolution equations, we show explicitly the connection between them for the specific case of the one-gluon energy distribution. 

The evolution equation for the one-gluon energy distribution, as derived from the generating functional, reads
\beq\label{evol1}
-\frac{\del}{\del \tau_0} D(x,\tau-\tau_0 )= \int^1_0 dz\, \hat{\cal K} (z)\, D\left({x\over z} ,{\tau-\tau_0 \over \sqrt{z}}\right) -\frac{1}{2}\int^1_0 dz \,\hat{\cal K}(z)\,D\left(x,\tau-\tau_0\right) \,.
\eeq
It is understood, here and in the following equation that $D(x>1)=0$, so that the lower bound on the first $z$-integration is actually $z=x$. 
To write down this equation we have assumed that the energy of the initial parton is $p_0^+$. It is convenient, for the foregoing derivation, to consider an arbitrary initial $p^+$, so that we shift $p_0^+$ to $x'p_0^+$. Under such a shift $\tau \rightarrow \tau/\sqrt{x'}$ (recall Eq.~(\ref{hatK})). We can then rewrite Eq. (\ref{evol1}) as 
\beq\label{evol2}
-\frac{\del}{\del \tau_0} D\left({x\over x'},{\tau-\tau_0 \over \sqrt{x'} } \right)= \frac{1}{\sqrt{x'}}\int^1_{0} dz \,\hat{\cal K} (z)\, D\left( {x\over zx'} ,{\tau-\tau_0 \over \sqrt{zx'} } \right) 
-\frac{1}{2\sqrt{x'}}\int^1_0 dz\, \hat{\cal K}(z)\,D\left({x\over x'},{\tau-\tau_0 \over \sqrt{x'} }\right).\nn
\eeq
Now let us introduce the following identity (which results from the Chapman-Kolmogorov law of composition of probabilities)
\beq\label{comp-ID}
D(x,\tau-\tau_0)= \int_x^1 \frac{dx' }{x'}\, D\left({x\over x'},{\tau-\tau' \over \sqrt{x'} }\right ) D(x',\tau'-\tau_0).
\eeq
This equality holds for any $\tau'$. In particular, it  is obviously true for $\tau'=\tau_0$ where $D(x',0)=\delta(x'-1)$, and for $\tau'=\tau$ where $D(x/x',0)=x\delta(x-x')$. More generally, taking the derivative of Eq.~(\ref{comp-ID}) with respect to $\tau'$ one gets\beq\label{comp-ID2}
-\int_x^1\frac{\rmd x'}{x'} \,\frac{\del}{\del \tau'} D\left({x\over x'},{\tau-\tau' \over \sqrt{x'} }\right ) D(x',\tau'-\tau_0)= \int_x^1\frac{\rmd x'}{x'} \,D\left({x\over x'},{\tau-\tau' \over \sqrt{x'} }\right ) \frac{\del}{\del \tau'} D(x',\tau'-\tau_0).\nn
\eeq
By combining this equation with Eq.~(\ref{evol2}) (in which we replace $\tau_0$ by $\tau'$)  we get
\beq\label{evol3}
\int_x^1\frac{\rmd x'}{x'}D\left({x\over x'},{\tau-\tau' \over \sqrt{x'} }\right ) \frac{\del}{\del \tau'} D(x',\tau'-\tau_0)&=&\int_x^1\frac{\rmd x'}{x'}\frac{1}{\sqrt{x'}}\int^1_{0} dz \hat{\cal K} (z)\, D\left( {x\over zx'} ,{\tau-\tau' \over \sqrt{zx'} } \right) D(x',\tau'-\tau_0)\nn
&-&\int_x^1\frac{\rmd x'}{x'} \frac{1}{2\sqrt{x'}}\int^1_0 dz \hat{\cal K}(z)\,D\left({x\over x'},{\tau-\tau' \over \sqrt{x'} }\right )D(x',\tau'-\tau_0)\,.\nn
\eeq
At this point we set $\tau'=\tau$, which allows us to perform the $x'$ integrations (thanks to the properties recalled after Eq.~(\ref{comp-ID})). We end up with 
\beq\label{evol5}
\frac{\del}{\del \tau} D(x,\tau-\tau_0)= \int^1_{x} dz\, \hat{\cal K} (z)\, \sqrt{\frac{z}{x}}\,D\left(\frac{x}{z},\tau-\tau_0\right)
- \frac{1}{2\sqrt{x}}\int^1_0 dz\, \hat{\cal K}(z)\,D(x,\tau-\tau_0).
\eeq
Since $\hat{\cal K}(z)$ is symmetric under the transformation $z\to1-z$, we have
\beq
\int_0^1dz\,z\hat{\cal K}(z)=\int_0^1dz\,(1-z)\hat{\cal K}(z)=\frac{1}{2}\int_0^1dz\,\hat{\cal K}(z)\,,
\eeq
which allows us to write Eq. (\ref{evol5}) as
\beq\label{evol7}
&&\frac{\del}{\del \tau} D(x,\tau)= \int^1_{x} dz\, \hat{\cal K} (z)\, \sqrt{\frac{z}{x}}\,D\left(\frac{x}{z},\tau\right)- \int^1_0 dz\,\frac{z}{\sqrt{x}} \hat{\cal K}(z)\,D(x,\tau),\nn
\eeq
which is the evolution equation (\ref{Dfin}).


\begin{thebibliography}{10}

\bibitem{Aad:2010bu}
{\bf Atlas Collaboration} Collaboration, G.~Aad {\em et.~al.}, {\it
  {Observation of a Centrality-Dependent Dijet Asymmetry in Lead-Lead
  Collisions at $\sqrt{s_{NN}}=2.77$ TeV with the ATLAS Detector at the LHC}},
  {\em Phys.Rev.Lett.} {\bf 105} (2010) 252303,
  [\href{http://xxx.lanl.gov/abs/1011.6182}{{\tt arXiv:1011.6182}}].

\bibitem{Chatrchyan:2011sx}
{\bf CMS Collaboration} Collaboration, S.~Chatrchyan {\em et.~al.}, {\it
  {Observation and studies of jet quenching in PbPb collisions at
  nucleon-nucleon center-of-mass energy = 2.76 TeV}},  {\em Phys.Rev.} {\bf
  C84} (2011) 024906, [\href{http://xxx.lanl.gov/abs/1102.1957}{{\tt
  arXiv:1102.1957}}].

\bibitem{Chatrchyan:2012nia}
{\bf CMS Collaboration} Collaboration, S.~Chatrchyan {\em et.~al.}, {\it {Jet
  momentum dependence of jet quenching in PbPb collisions at
  $\sqrt{s_{NN}}=2.76$ TeV}},  {\em Phys.Lett.} {\bf B712} (2012) 176--197,
  [\href{http://xxx.lanl.gov/abs/1202.5022}{{\tt arXiv:1202.5022}}].

\bibitem{Aad:2012vca}
{\bf ATLAS Collaboration} Collaboration, G.~Aad {\em et.~al.}, {\it
  {Measurement of the jet radius and transverse momentum dependence of
  inclusive jet suppression in lead-lead collisions at $\sqrt{s_{NN}}$= 2.76
  TeV with the ATLAS detector}},  {\em Phys.Lett.} {\bf B719} (2013) 220--241,
  [\href{http://xxx.lanl.gov/abs/1208.1967}{{\tt arXiv:1208.1967}}].

\bibitem{Chatrchyan:2013kwa}
{\bf CMS Collaboration} Collaboration, S.~Chatrchyan {\em et.~al.}, {\it
  {Modification of jet shapes in PbPb collisions at $\sqrt{s_{NN}}$ = 2.76
  TeV}},  \href{http://xxx.lanl.gov/abs/1310.0878}{{\tt arXiv:1310.0878}}.

\bibitem{Baier:1996kr}
R.~Baier, Y.~L. Dokshitzer, A.~H. Mueller, S.~Peigne, and D.~Schiff, {\it
  {Radiative energy loss of high-energy quarks and gluons in a finite volume
  quark - gluon plasma}},  {\em Nucl.Phys.} {\bf B483} (1997) 291--320.

\bibitem{Baier:1996sk}
R.~Baier, Y.~L. Dokshitzer, A.~H. Mueller, S.~Peigne, and D.~Schiff, {\it
  {Radiative energy loss and p(T) broadening of high-energy partons in
  nuclei}},  {\em Nucl.Phys.} {\bf B484} (1997) 265--282.

\bibitem{Baier:1998kq}
R.~Baier, Y.~L. Dokshitzer, A.~H. Mueller, and D.~Schiff, {\it {Medium induced
  radiative energy loss: Equivalence between the BDMPS and Zakharov
  formalisms}},  {\em Nucl.Phys.} {\bf B531} (1998) 403--425,
  [\href{http://xxx.lanl.gov/abs/hep-ph/9804212}{{\tt hep-ph/9804212}}].

\bibitem{Zakharov:1996fv}
B.~Zakharov, {\it {Fully quantum treatment of the Landau-Pomeranchuk-Migdal
  effect in QED and QCD}},  {\em JETP Lett.} {\bf 63} (1996) 952--957.

\bibitem{Zakharov:1997uu}
B.~Zakharov, {\it {Radiative energy loss of high-energy quarks in finite size
  nuclear matter and quark - gluon plasma}},  {\em JETP Lett.} {\bf 65} (1997)
  615--620.

\bibitem{Wiedemann:2000za}
U.~A. Wiedemann, {\it {Gluon radiation off hard quarks in a nuclear
  environment: Opacity expansion}},  {\em Nucl.Phys.} {\bf B588} (2000)
  303--344, [\href{http://xxx.lanl.gov/abs/hep-ph/0005129}{{\tt
  hep-ph/0005129}}].

\bibitem{Gyulassy:2000er}
M.~Gyulassy, P.~Levai, and I.~Vitev, {\it {Reaction operator approach to
  nonAbelian energy loss}},  {\em Nucl.Phys.} {\bf B594} (2001) 371--419,
  [\href{http://xxx.lanl.gov/abs/nucl-th/0006010}{{\tt nucl-th/0006010}}].

\bibitem{Arnold:2002ja}
P.~B. Arnold, G.~D. Moore, and L.~G. Yaffe, {\it {Photon and gluon emission in
  relativistic plasmas}},  {\em JHEP} {\bf 0206} (2002) 030,
  [\href{http://xxx.lanl.gov/abs/hep-ph/0204343}{{\tt hep-ph/0204343}}].

\bibitem{MehtarTani:2010ma}
Y.~Mehtar-Tani, C.~A. Salgado, and K.~Tywoniuk, {\it {Anti-angular ordering of
  gluon radiation in QCD media}},  {\em Phys.Rev.Lett.} {\bf 106} (2011)
  122002, [\href{http://xxx.lanl.gov/abs/1009.2965}{{\tt arXiv:1009.2965}}].

\bibitem{MehtarTani:2011tz}
Y.~Mehtar-Tani, C.~Salgado, and K.~Tywoniuk, {\it {Jets in QCD Media: From
  Color Coherence to Decoherence}},  {\em Phys.Lett.} {\bf B707} (2012)
  156--159, [\href{http://xxx.lanl.gov/abs/1102.4317}{{\tt arXiv:1102.4317}}].

\bibitem{CasalderreySolana:2011rz}
J.~Casalderrey-Solana and E.~Iancu, {\it {Interference effects in
  medium-induced gluon radiation}},  {\em JHEP} {\bf 1108} (2011) 015,
  [\href{http://xxx.lanl.gov/abs/1105.1760}{{\tt arXiv:1105.1760}}].

\bibitem{MehtarTani:2011gf}
Y.~Mehtar-Tani, C.~A. Salgado, and K.~Tywoniuk, {\it {The radiation pattern of
  a QCD antenna in a dilute medium}},  {\em JHEP} {\bf 1204} (2012) 064,
  [\href{http://xxx.lanl.gov/abs/1112.5031}{{\tt arXiv:1112.5031}}].

\bibitem{MehtarTani:2011jw}
Y.~Mehtar-Tani and K.~Tywoniuk, {\it {Jet coherence in QCD media: the antenna
  radiation spectrum}},  {\em JHEP} {\bf 1301} (2013) 031,
  [\href{http://xxx.lanl.gov/abs/1105.1346}{{\tt arXiv:1105.1346}}].

\bibitem{CasalderreySolana:2012ef}
J.~Casalderrey-Solana, Y.~Mehtar-Tani, C.~A. Salgado, and K.~Tywoniuk, {\it
  {New picture of jet quenching dictated by color coherence}},  {\em
  Phys.Lett.} {\bf B725} (2013) 357--360,
  [\href{http://xxx.lanl.gov/abs/1210.7765}{{\tt arXiv:1210.7765}}].

\bibitem{Blaizot:2012fh}
J.-P. Blaizot, F.~Dominguez, E.~Iancu, and Y.~Mehtar-Tani, {\it {Medium-induced
  gluon branching}},  {\em JHEP} {\bf 1301} (2013) 143,
  [\href{http://xxx.lanl.gov/abs/1209.4585}{{\tt arXiv:1209.4585}}].

\bibitem{Blaizot:2013hx}
J.-P. Blaizot, E.~Iancu, and Y.~Mehtar-Tani, {\it {Medium-induced QCD cascade:
  democratic branching and wave turbulence}},  {\em Phys.Rev.Lett.} {\bf 111}
  (2013) 052001, [\href{http://xxx.lanl.gov/abs/1301.6102}{{\tt
  arXiv:1301.6102}}].

\bibitem{Liou:2013qya}
T.~Liou, A.~Mueller, and B.~Wu, {\it {Radiative $p_\bot$-broadening of
  high-energy quarks and gluons in QCD matter}},  {\em Nucl.Phys.} {\bf A916}
  (2013) 102--125, [\href{http://xxx.lanl.gov/abs/1304.7677}{{\tt
  arXiv:1304.7677}}].

\bibitem{Altarelli:1977zs}
G.~Altarelli and G.~Parisi, {\it {Asymptotic Freedom in Parton Language}},
  {\em Nucl.Phys.} {\bf B126} (1977) 298.

\bibitem{Cvitanovic:1980ru}
P.~Cvitanovic, P.~Hoyer, and K.~Zalewski, {\it {Parton evolution as a branching
  process}},  {\em Nucl.Phys.} {\bf B176} (1980) 429.

\bibitem{Baier:2000sb}
R.~Baier, A.~H. Mueller, D.~Schiff, and D.~Son, {\it {'Bottom up'
  thermalization in heavy ion collisions}},  {\em Phys.Lett.} {\bf B502} (2001)
  51--58, [\href{http://xxx.lanl.gov/abs/hep-ph/0009237}{{\tt
  hep-ph/0009237}}].

\bibitem{Jeon:2003gi}
S.~Jeon and G.~D. Moore, {\it {Energy loss of leading partons in a thermal QCD
  medium}},  {\em Phys.Rev.} {\bf C71} (2005) 034901,
  [\href{http://xxx.lanl.gov/abs/hep-ph/0309332}{{\tt hep-ph/0309332}}].

\bibitem{Schenke:2009gb}
B.~Schenke, C.~Gale, and S.~Jeon, {\it {MARTINI: An Event generator for
  relativistic heavy-ion collisions}},  {\em Phys.Rev.} {\bf C80} (2009)
  054913, [\href{http://xxx.lanl.gov/abs/0909.2037}{{\tt arXiv:0909.2037}}].

\end{thebibliography}



\end{document}